\documentclass{PoS}

\renewcommand{\bar}[1]{\overline{#1}}

\newcommand{\ket}[1]{\,\left|\,{#1}\right\rangle}
\newcommand{\mbf}{\mathbf}

\title{Novel High Transverse Momentum Phenomena in Hadronic and Nuclear Collisions}

\ShortTitle{Novel High Transverse Momentum Phenomena}

\author{\speaker{Stanley J. Brodsky}\thanks{SLAC-PUB 13571. This work was supported in
part by the Department of Energy, contract No. DE-AC02-76SF00515.}\\
SLAC National Accelerator Laboratory, Stanford University, Stanford, California 94309\\
E-mail:sjbth@slac.stanford.edu}

\abstract{I discuss a number of novel phenomenological features of QCD in high transverse momentum reactions. The
presence of direct higher-twist processes, where a proton is produced directly in the hard subprocess, can explain the ``baryon anomaly" --  the large proton-to-pion ratio seen at RHIC in
high centrality heavy ion collisions.  Direct hadronic processes can also account for the deviation from leading-twist PQCD scaling at fixed $x_T= 2 p_T/\sqrt s.$    I suggest that the ``ridge"  -- the same-side  long-range rapidity correlation observed at RHIC in high centrality  heavy ion collisions is due to the imprint  of semihard DGLAP gluon radiation from initial-state partons which have transverse momenta biased toward the trigger.  A model for early thermalization of the quark-gluon medium is also outlined.   Rescattering interactions from gluon-exchange, normally neglected in the parton model, have a profound effect in QCD hard-scattering reactions,
leading to leading-twist single-spin asymmetries, diffractive deep inelastic scattering, diffractive hard hadronic reactions, the breakdown of
the Lam-Tung relation in Drell-Yan reactions, nuclear shadowing ---all leading-twist dynamics not incorporated in
the light-front wavefunctions of the target computed in isolation.  Antishadowing is shown to be quark flavor  specific and thus different in charged and neutral deep inelastic lepton-nucleus scattering.  I also discuss other aspects of quantum effects in heavy ion collisions, such as tests of hidden color in nuclear wavefunctions, the use of
diffraction to materialize the Fock states of a hadronic projectile and test QCD color transparency, and the important consequences of color-octet intrinsic heavy quark distributions in the proton for particle and Higgs production at high $x_F$.  I also discuss how the AdS/CFT correspondence between Anti-de Sitter space and conformal
gauge theories allows one to compute the
analytic form of  frame-independent light-front wavefunctions of mesons and baryons and to compute quark and gluon hadronization at the amplitude level.  Finally,  the BLM method for determining  the renormalization scale in PQCD calculations is  reviewed.
}

\FullConference{High-$p_T$ Physics at LHC-09\\
		 February 4 - 7, 2009\\
		 Prague, Czech Republic}

\begin{document}

\section{Introduction}

\begin{figure}[!]
 \begin{center}
\includegraphics[width=15.0cm]{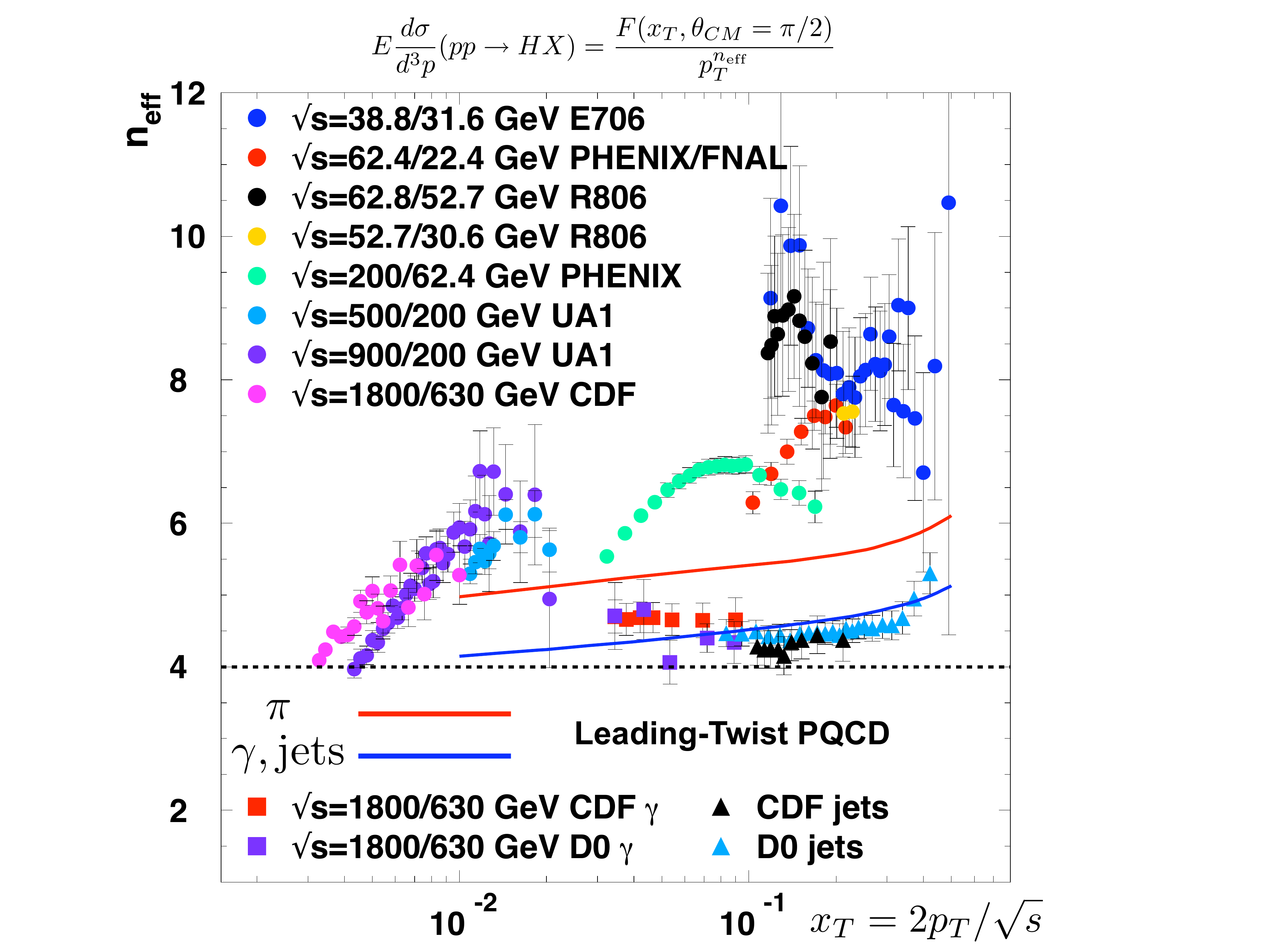}
\end{center}
  \caption{Comparison of RHIC and fixed-target data for hadron, isolated photon, and jet production with the leading-twist PQCD predictions for the power-falloff of  the semi-inclusive cross section 
  $E {d \sigma/ d^3p}(p p \to H X) = {F(x_T, \theta_{CM}=\pi/2)/p_T^{n_{\rm eff}}}$  at fixed $x_T$.  
The data from R806, PHENIX, ISR/FNAL, E706 are for 
charged or neutral pion production, whereas the CDF, UA1 data at small $x_T$  are for charged 
hadrons. The blue curve is the prediction of  leading-twist QCD for isolated photon and jet production, including the scale-breaking effects of the  running coupling and evolution of the proton structure functions. The red curve is the QCD prediction for pion production, which also  includes the effect from the evolution of the fragmentation function. The dashed line at $n_{\rm eff} = 4$ is the prediction of the scale-invariant parton model.   From Arleo, et al.~\cite{Arleo}. }
\label{figNew1B}  
\end{figure}

\begin{figure}[htb]
\centering
\includegraphics[width=15.0cm]{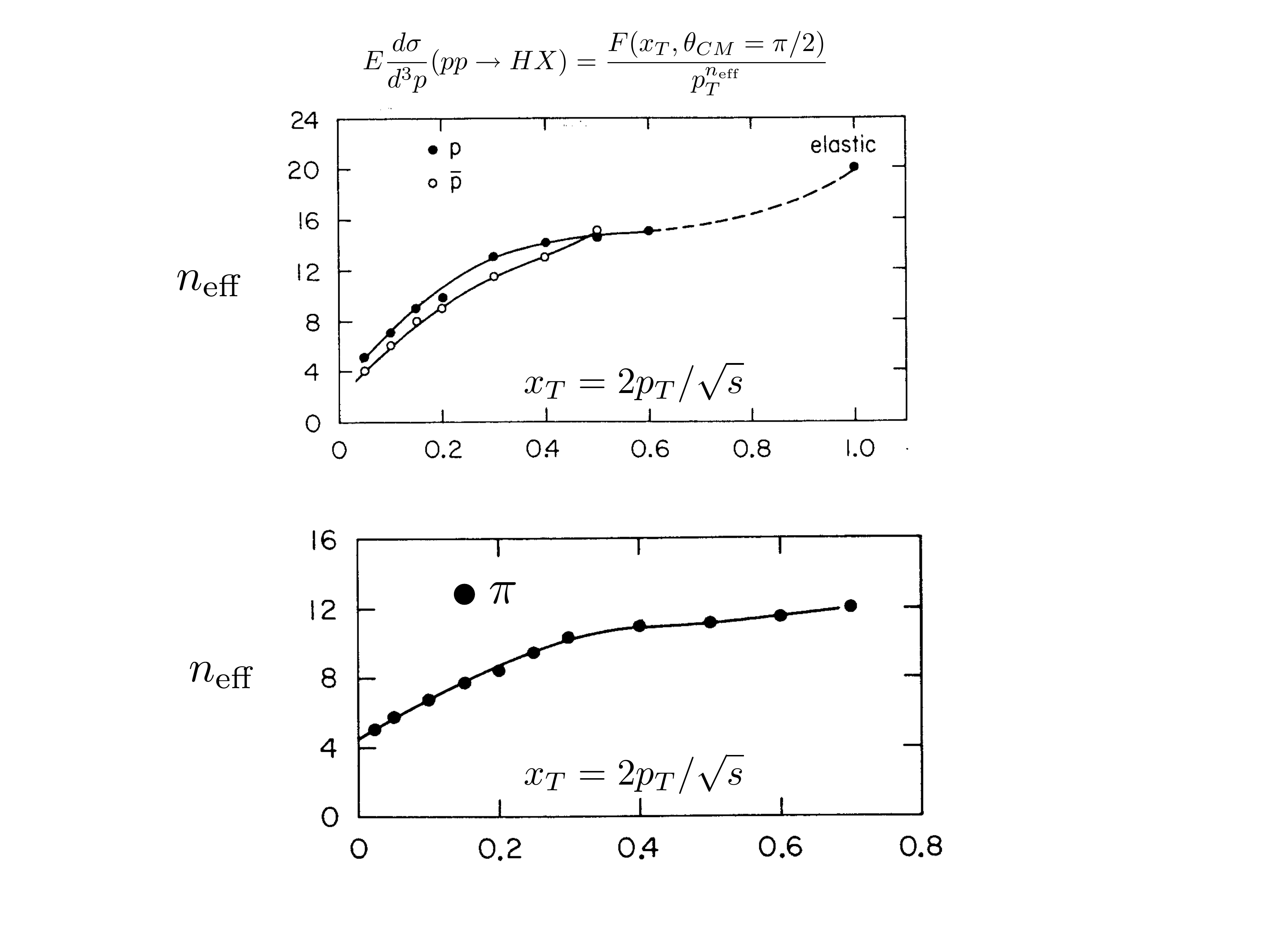}
\caption{Power-law scaling for particle production at large transverse momentum at fixed $x_T$  from the Chicago-Princeton fixed target experiments at FermiLab.~\cite{Cronin:1973fd,Antreasyan:1978cw}. } \label{figNew4}
 \end{figure}

The advent of the LHC will open up a new domain for testing QCD at energies up to $\sqrt s = 14$ TeV  in both proton-proton  and nucleus-nucleus collisions.      Measurements of particle production at high transverse momentum at RHIC and the Tevatron suggest that hard hadronic physics at the LHC may have surprising features.  For example, the  leading-twist pQCD prediction for high transverse momentum inclusive cross sections is obtained from the convolution of nearly scale-invariant two-to-two quark and gluon hard subprocess cross sections with the logarithmically evolved structure functions and fragmentation functions.  However,  as shown in fig.\ref {figNew1B},   measurements of the inclusive cross section for hadron production at fixed $x_T= 2 p_T/\sqrt x$ and fixed $\theta_{CM}$ for a wide range of high energy kinematics do not show the near-conformal scaling expected from  leading-twist QCD. The anomalous fixed-$x_T$ scaling at high energies is similar to that found for meson and baryon production at lower energies by the Chicago-Princeton fixed-target experiment at FermiLab,~\cite{Cronin:1973fd}, as shown in fig.\ref{figNew4}.    In contrast, the scaling of the measured inclusive cross sections for both isolated direct photon production and jet triggers as shown in fig.\ref {figNew1B} are both in good agreement with perturbative QCD.

\begin{figure}[!]
 \begin{center}
\includegraphics[width=18.0cm]{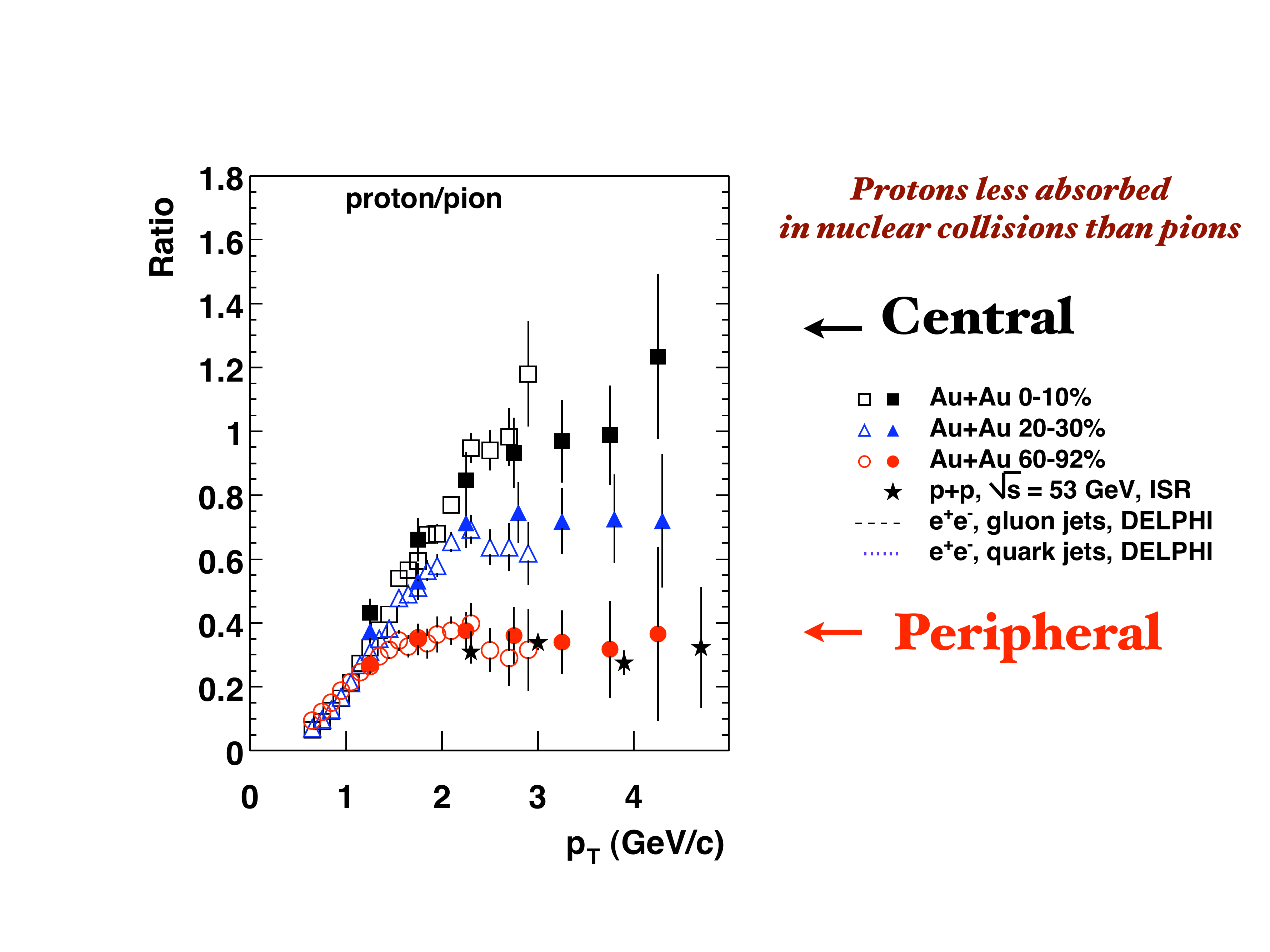}
\end{center}
 \caption{The ratio of protons to pions produced at large $p_T$ in heavy ion collisions as a function of centrality from the PHENIX experiment at
RHIC~\cite{Adler:2003kg}.  The open and solid symbols indicate neutral and charged pions, respectively.   A comparison with the measured $p/\pi$ ratio in $e^+
e^-$ and $ p p $ reactions is also shown.
The anomalous rise of the $p /\pi$ ratio with $p_T$ at high centrality is
consistent with the hypothesis that only the pions are absorbed in the nuclear medium.}
\label{figNew3}  
\end{figure}

\begin{figure}[!]
 \begin{center}
\includegraphics[width=15.0cm]{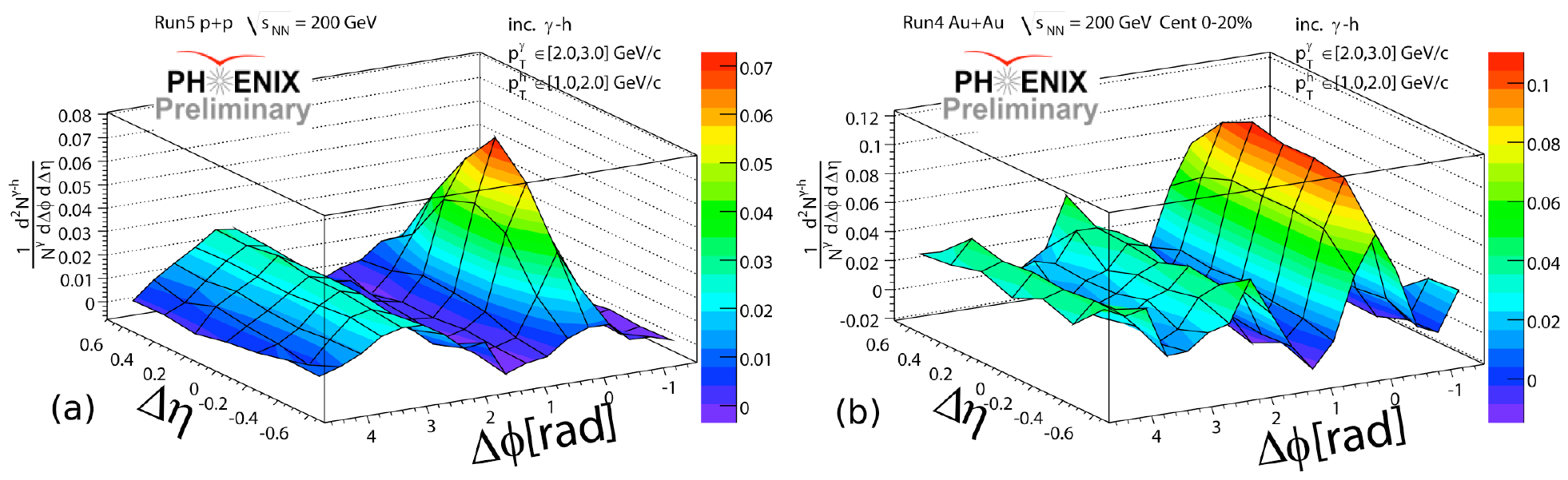}
\end{center}
 \caption{The ``ridge"  -- the same-side long range rapidity correlation associated with a high $p_T$ photon which comes predominately from $\pi^0$ decay. Comparison of  proton-nucleus  (a ) and nucleus-nucleus  (b) collisions. From the PHENIX collaboration, ref.~\cite{McCumber:2008id} }.
\label{figNew3A}  
\end{figure}

Recent experimental results from heavy ion collisions at RHIC have established remarkable and unexpected nuclear phenomena.  One observes the  ``baryon anomaly~\cite{Adler:2003kg}: an increasing baryon-to-meson ratio with centrality (see fig.\ref{figNew3}), just opposite to the conventional expectation that protons should suffer more energy loss in the nuclear medium than mesons.  Remarkably, one also sees long-range rapidity correlations (the ``ridge")  in association with a high $p_T$ hadron trigger~\cite{McCumber:2008id}  (see fig.\ref{figNew3A}).   Furthermore, the ridge  disappears in the case of dihadron triggers which balance in transverse momenta~\cite{Olga} .

In this talk I will  discuss QCD mechanisms which could account  for the baryon anomaly, the violation  of leading twist scaling, and the ridge phenomena.  For example, the
presence of color-transparent direct higher-twist processes, where a proton is produced directly in the hard subprocess, can explain the large proton-to-pion ratio seen in
high centrality heavy ion collisions as well as the anomalous scaling of the inclusive cross section at fixed $x_T.$  I also argue that the ridge is a consequence of the semihard gluons associated with the trigger-biased DLGLAP evolution of the initial-state quark and gluon distributions,  imprinted on the nuclear medium.

Initial- and
final-state interactions from gluon-exchange, normally neglected in the parton model, have important effects in QCD hard-scattering reactions,
leading to leading-twist single-spin asymmetries, diffractive deep inelastic scattering, diffractive hard hadronic reactions, the breakdown of
the Lam Tung relation in Drell-Yan reactions, as well as nuclear shadowing and non-universal antishadowing -- leading-twist physics not incorporated in
the light-front wavefunctions of the target computed in isolation.
Thus the standard factorization picture is oversimplified due to nontrivial gluonic rescattering.  Furthermore, as I will discuss,  even the nuclear modifications of structure functions are not universal, but depend on the 	quark flavor.

I will also discuss other aspects of quantum effects in heavy ion collisions, such as tests of hidden color in nuclear wavefunctions, the use of
diffraction to materialize the Fock states of a hadronic projectile and test color transparency, and the consequences of color-octet  intrinsic heavy quarks such as the strong nuclear suppression of $J/\psi$ hadroproduction at high $x_F.$   A model for early thermalization of the quark-gluon medium is presented.  I also will review how the AdS/CFT correspondence between Anti-de Sitter space and conformal
gauge theories provides an exact correspondence which allows one to compute the
analytic form of the frame-independent light-front wavefunctions of mesons and baryons, thus providing a method to compute hadronization at the amplitude level.  Finally, I will discuss the elimination of the renormalization scale ambiguity in PQCD calculations.

\section{Dynamical Higher-Twist Processes and Fixed-$x_T$ scaling}

As noted in the introduction, the fundamental test of leading-twist QCD predictions in high transverse momentum hadronic reactions is the measurement of the power-law
fall-off of the inclusive cross section 
${E d \sigma/d^3p}(A B \to C X) ={ F(\theta_{cm}, x_T)/ p_T^{n_eff} } $ at fixed $x_T = 2 p_T/\sqrt s$ 
and fixed $\theta_{CM},$ where $n_{eff} \sim 4 + \delta$. Here $\delta  \le 1$ is the correction to the conformal prediction arising
from the QCD running coupling and DGLAP evolution of the input distribution and fragmentation functions~\cite{Brodsky:2005fz,Arleo}.  
The usual expectation is that leading-twist subprocesses will dominate measurements of high $p_T$ hadron production at RHIC and Tevatron energies. Indeed, the  data for isolated photon production $ p p \to \gamma_{\rm direct} X$ as well as jet production agrees well with the  leading-twist scaling prediction $n_{eff}  \simeq 4.5$ as seen in fig.\ref{figNew1B} ~\cite{Arleo}. 
However, as seen in fig.\ref{figNew1B},  measurements  of  $n_{eff} $ for $ p p \to \pi X$  are not consistent with the leading twist predictions.  
Striking
deviations from the leading-twist predictions were also observed at lower energy at the ISR and  Fermilab fixed-target experiments~\cite{Sivers:1975dg,Cronin:1973fd,Antreasyan:1978cw}.  
See fig.\ref{figNew4}. The high values $n_{eff}$ with $x_T$ seen in the data  indicate the presence of an array of higher-twist processes, including subprocesses where the hadron enters directly, rather than through jet fragmentation~\cite{Blankenbecler:1975ct}.

It should be emphasized that the existence of dynamical higher-twist processes in which a hadron interacts directly within a hard subprocess is a prediction of
QCD.   For example, the subprocess $\gamma^* q \to \pi q,$ where the pion is produced directly through the pion's $\bar q q \to \pi$ distribution amplitude $\phi_\pi(x,Q)$ underlies  deeply virtual meson scattering $\gamma p \to \pi X.$ 
The corresponding timelike subprocess  $\pi q \to \gamma^*q$ dominates the Drell-Yan reaction $\pi p \to \ell^+ \ell^- X $ at high $x_F$ ~\cite{Berger:1979du}, thus accounting for the change in angular distribution from the canonical $1+ \cos^2\theta$ distribution for transversely polarized virtual photons to $\sin^2\theta$ corresponding to longitudinal photons;  the virtual photon becomes longitudinally polarized  at
high $x_F$, reflecting  the spin of the pion entering the QCD hard subprocess.
Crossing predicts 
reactions where the final-state hadron appears directly in the subprocess such as $e^+ e^- \to \pi X$ at $z=1$.
The nominal power-law fall-off at  fixed $x_T$ is set by the number of elementary fields entering the hard subprocess $n_{\rm eff} = 2 n_{\rm active} -4.$   The power-law fall-off $(1-x_T)^F$ at high $x_T$ is set by the total number of spectators  $F = 2 n_{\rm spectators} -1 ~\cite{Blankenbecler:1975ct}.$ 

It should also be noted that direct higher-twist subprocesses, where the trigger hadron is produced  within the hard subprocess avoid the waste of same-side energy, thus allowing the target and projectile structure functions to be evaluated at the minimum values of $x_1$ and $x_2$ where they are at their maximum.  

Examples of direct baryon and meson higher-twist subprocesses are: $u d \to \Lambda \bar s , u \bar d \to \pi^+ g, u g \to \pi^+ d, u \bar s \to K^+ g, u g \to K^+ s.$
These direct subprocesses involve the distribution amplitude of the hadron which has dimension  $\Lambda_{QCD}$ for mesons and
$\Lambda^2_{QCD}$ for baryons; thus these higher-twist contributions to the inclusive cross section $Ed \sigma/ d^3p$ at fixed $x_T$
nominally scale as $\Lambda^2_{QCD}/p_T^6$ for mesons and $\Lambda^4_{QCD}/p_T^8$ for baryons.

Note  also that at $x_T\to 1$,  the inclusive reaction $ p p \to p X$ must conform by duality with the scaling  of fixed $\theta_{cm}$ elastic $ p p \to p p $ scattering at large momentum transfer where $ n \simeq 20$ as illustrated in fig.\ref{figNew4}.  
It should also be noted that the DGLAP evolution of structure functions is derived in PQCD for free quarks and gluons.  However at large $x \to 1$ and $z \to 1$  a bound state  quark or gluon becomes far-off shell, thus quenching the radiative processes. This analytic modification allows one to have continuity between hard exclusive reactions and inclusive reactions in PQCD.

The behavior of the single-particle inclusive cross section will be a key test of QCD at the LHC, since the leading-twist prediction for $n_{\rm eff} \sim 4 + \delta$ is independent of the detailed form of the structure and fragmentation functions.

\section{Direct Higher-Twist Processes and the Baryon Anomaly}

\begin{figure}[!]
 \begin{center}
\includegraphics[width=15.0cm]{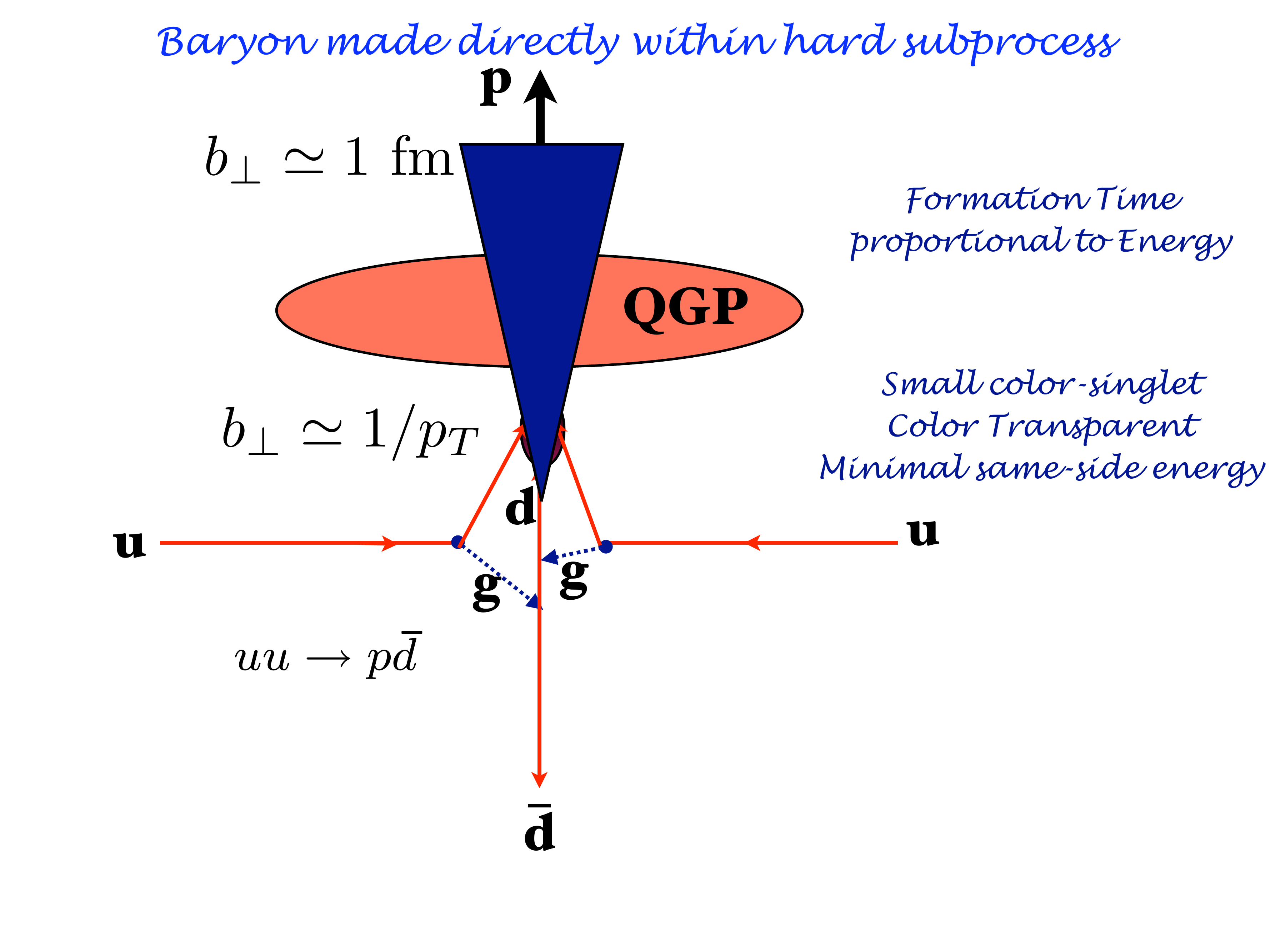}
\end{center}
  \caption{Mechanism for Higher-Twist Direct Proton Production. The small-transverse size color-singlet 3-quark state is not affected by the nuclear background.}
\label{figNew2}  
\end{figure}

The fixed
$x_T$ scaling of the  proton production cross section  ${E d \sigma/ d^3p}(p p \to p p X)$ is particularly anomalous,  far from the $1/p^4_T$ to $1/p^5_T$ scaling predicted by pQCD~\cite{Brodsky:2005fz}.   See  fig.\ref{figNew4}.   Sickles and I have argued that this anomalous scaling is due to subprocesses~\cite{Brodsky:2005fz} , where the
proton is  created directly within the hard reaction, such as   $ u u \to p \bar d$, as illustrated in fig.\ref{figNew2}.
The fragmentation of a gluon or quark jet
to a proton requires that the underlying  2 to 2 subprocess occurs at a  higher transverse momentum than the $p_T$ of the observed proton because of the fast-falling $(1-z)^3 $ quark-to-proton fragmentation function; in contrast,  the direct subprocess is maximally energy efficient.   Such ``direct" reactions can readily explain the fast-falling power-law  falloff
observed at fixed $x_T$ and fixed-$\theta_{cm}$ at the ISR, FermiLab and RHIC~\cite{Brodsky:2005fz}.

\begin{figure}[htb]
\centering
\includegraphics[width=15.0cm]{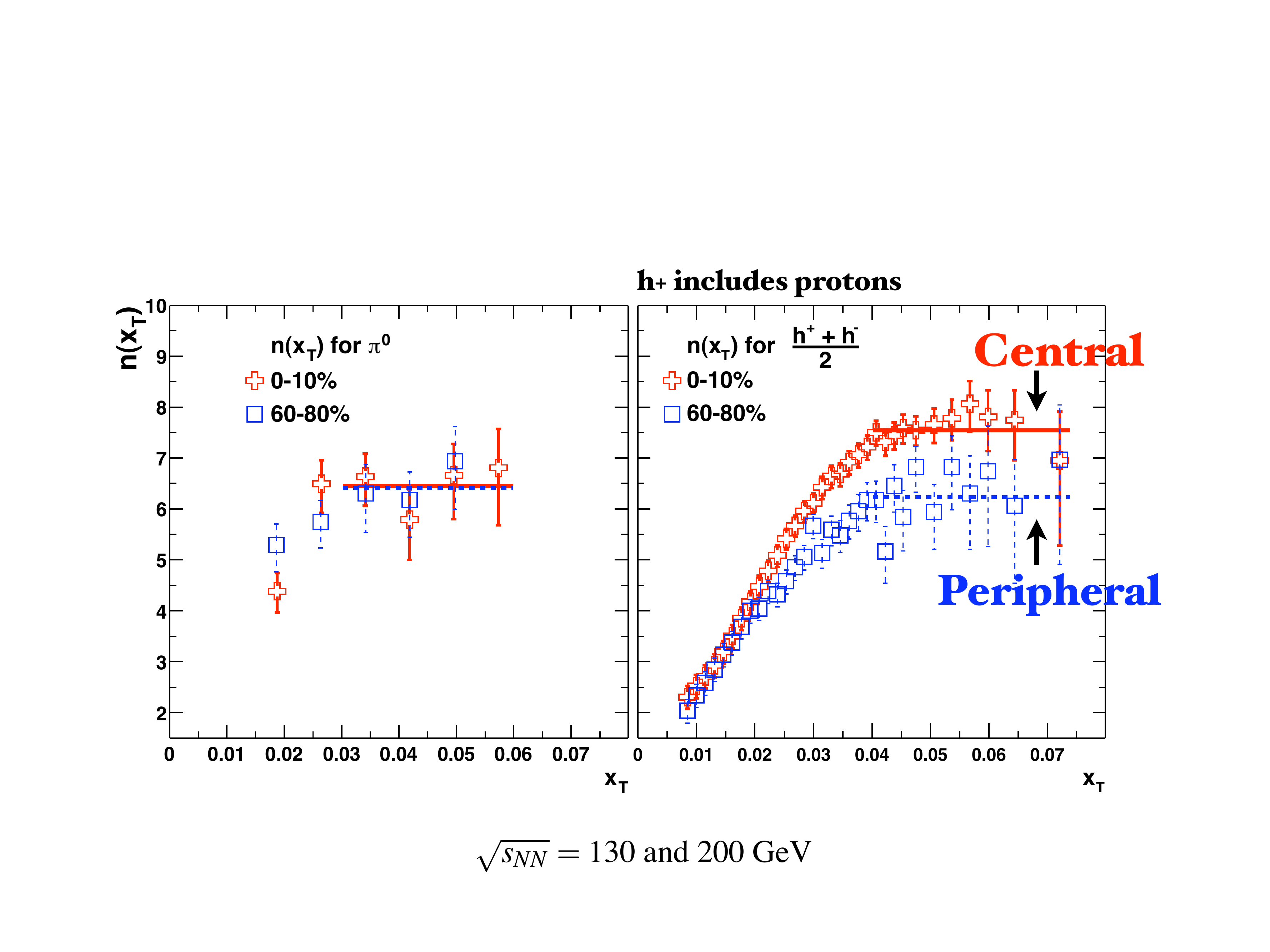}
\caption{The power-law scaling index $n_{eff}$ at fixed $x_T$  as a function of  centrality. The positive charged hadron trigger is dominated by protons at high $p_T$ for central collisions, consistent with the color transparency of direct higher-twist  baryon production processes. }
 \label{figNew6}
 \end{figure}
 
 \begin{figure}[htb]
\centering
\includegraphics[width=15.0cm]{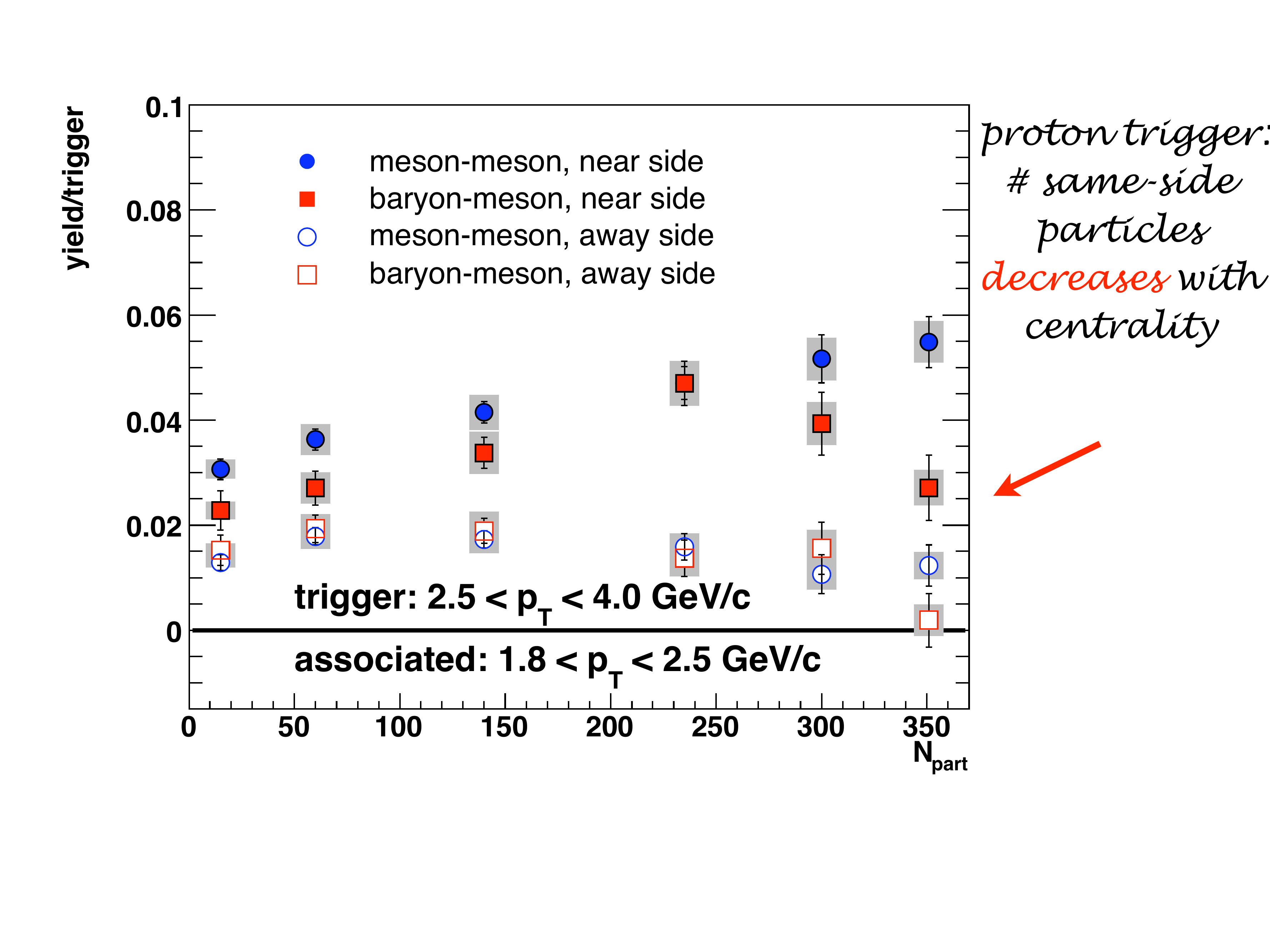}
\caption{Yield of same-side and away-side hadrons as a function of total multiplicity.  One sees a decrease in same-side yield for proton triggers at high centrality, consistent with the onset of direct proton production. }
 \label{figNew6A}
 \end{figure}

 \begin{figure}[htb]
\centering
\includegraphics[width=15.0cm]{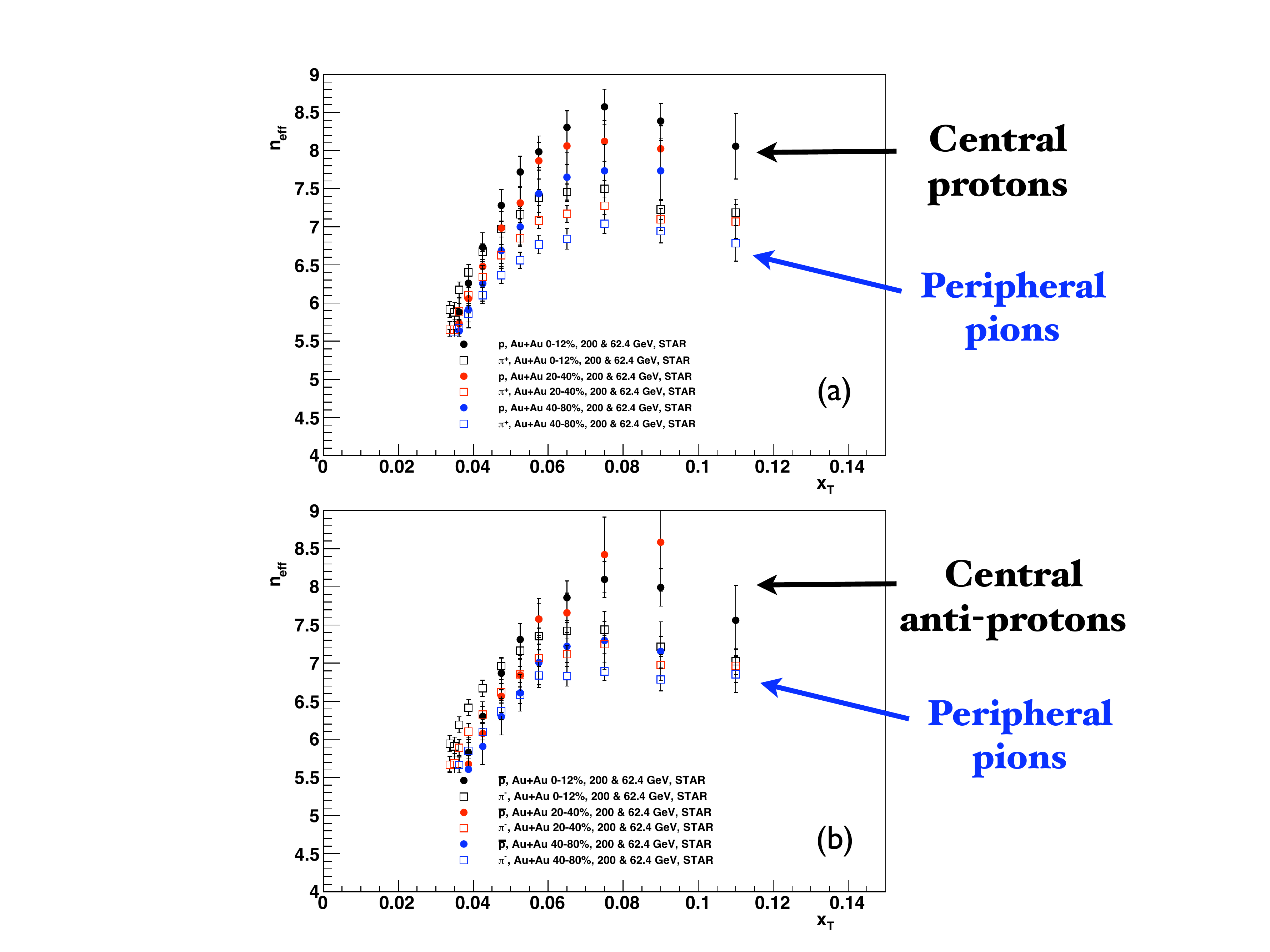}
\caption{The power-law scaling index $n_{eff}$ at fixed $x_T$  as a function 
of  centrality ~\cite{SicklesPrivate}    (a) Positively charged hadron triggers. (b) Negatively charged hadron triggers. The increase of $n_{eff}$ with centrality is consistent with the nuclear survival of direct higher-twist subprocesses. }
 \label{figNew6C}
 \end{figure}

Since the proton  is initially produced as a small-size $b_\perp \sim 1/p_T$  color-singlet state, it is ``color transparent" ~\cite{Brodsky:1988xz}   and can thus propagate through dense nuclear matter with minimal energy loss.  In
contrast, the pions which are  produced from jet fragmentation have a  normal inelastic cross section. This provides a plausible explanation of the RHIC
data~\cite{Adler:2003kg}, which shows a dramatic rise of the $p \to \pi$ ratio at with increasing $p_T$ when one compares  peripheral with central heavy ion collisions, as illustrated in fig.\ref{figNew2}.  
The color transparency of the proton produced in the direct process also explains why the index $n_{eff}$ rises with centrality, as seen in fig.\ref{figNew6},  -- the higher-twist color-transparent subprocess dominates in the nuclear medium.   In addition,  the fact that the proton tends to be produced alone in a direct subprocess explains  why the yield of same-side hadrons along the proton trigger is diminished with increasing centrality.   
See  fig.\ref {figNew6A}.

Thus the QCD color transparency of directly produced baryons can  explain the baryon anomaly seen in heavy-ion collisions at RHIC: the color-transparent proton state is not absorbed, but a pion produced from fragmentation is
diminished in the nuclear medium. This pattern is confirmed in the compilation shown in fig.\ref{figNew6C} ~\cite{SicklesPrivate}. The increase of $n_{eff}$ with centrality is consistent with the nuclear survival of direct higher-twist subprocesses for both protons and antiprotons, and to a lesser extent, for mesons.

\section{Diffractive Deep Inelastic Scattering}
A remarkable feature of deep inelastic lepton-proton scattering at HERA is that approximately 10\% events are
diffractive~\cite{Adloff:1997sc,Breitweg:1998gc}: the target proton remains intact, and there is a large rapidity gap between the proton and the
other hadrons in the final state.  The presence of a rapidity gap
between the target and diffractive system requires that the target
remnant emerges in a color-singlet state; this is made possible in
any gauge by soft rescattering. 
This phenomenon can be understood as the fact that rescattering of the struck quark from gluon interactions with the target quark spectators in DIS is not suppressed -- the vector gluon neutralizes color in the t-channel, and integration over near-on-shell propagator produces the phase $i$ characteristic of Pomeron exchange.  The multiple scattering of the struck
parton via instantaneous interactions in the target generates
dominantly imaginary diffractive amplitudes, giving rise to an
effective ``hard pomeron'' exchange.

These diffractive deep inelastic scattering (DDIS) events can also be understood most simply from the perspective
of the color-dipole model: the $q \bar q$ Fock state of the high-energy virtual photon diffractively dissociates into a diffractive dijet
system.  The exchange of multiple gluons between  the color dipole of the $q \bar q$ and the quarks of the target proton neutralizes the color
separation and leads to the diffractive final state.  The same multiple gluon exchange also controls diffractive vector meson electroproduction
at large photon virtuality \cite{Brodsky:1994kf}.  This observation presents a paradox: if one chooses the conventional parton model frame where
the photon light-front momentum is negative $q+ = q^0 + q^z  < 0$, the virtual photon interacts with a quark constituent with light-cone
momentum fraction $x = {k^+/p^+} = x_{bj} . ~$   Furthermore, the gauge link associated with the struck quark (the Wilson line) becomes unity in
light-cone gauge $A^+=0$. Thus the struck ``current" quark apparently experiences no final-state interactions. Since the light-front
wavefunctions $\psi_n(x_i,k_{\perp i})$ of a stable hadron are real, it appears impossible to generate the required imaginary phase associated
with pomeron exchange, let alone large rapidity gaps.

This paradox was resolved by Hoyer, Marchal,  Peigne, Sannino and myself~\cite{Brodsky:2002ue}.  Consider the case where the virtual photon
interacts with a strange quark---the $s \bar s$ pair is assumed to be produced in the target by gluon splitting.  In the case of Feynman gauge,
the struck $s$ quark continues to interact in the final state via gluon exchange as described by the Wilson line. The final-state interactions
occur at a light-cone time $\Delta\tau \simeq 1/\nu$ shortly after the virtual photon interacts with the struck quark. When one integrates over
the nearly-on-shell intermediate state, the amplitude acquires an imaginary part. Thus the rescattering of the quark produces a separated
color-singlet $s \bar s$ and an imaginary phase. In the case of the light-cone gauge $A^+ = \eta \cdot A =0$, one must also consider the
final-state interactions of the (unstruck) $\bar s$ quark. The gluon propagator in light-cone gauge $d_{LC}^{\mu\nu}(k) = (i/k^2+ i
\epsilon)\left[-g^{\mu\nu}+\left(\eta^\mu k^\nu+ k^\mu\eta^\nu / \eta\cdot k\right)\right] $ is singular at $k^+ = \eta\cdot k = 0.$ The
momentum of the exchanged gluon $k^+$ is of ${ \cal O}{(1/\nu)}$; thus rescattering contributes at leading twist even in light-cone gauge. The
net result is  gauge invariant and is identical to the color dipole model calculation. The calculation of the rescattering effects on DIS in
Feynman and light-cone gauge through three loops is given in detail for an Abelian model in reference~\cite{Brodsky:2002ue}.  The result shows
that the rescattering corrections reduce the magnitude of the DIS cross section in analogy to nuclear shadowing.

\begin{figure}[!]
 \begin{center}
\includegraphics[width=15.0cm]{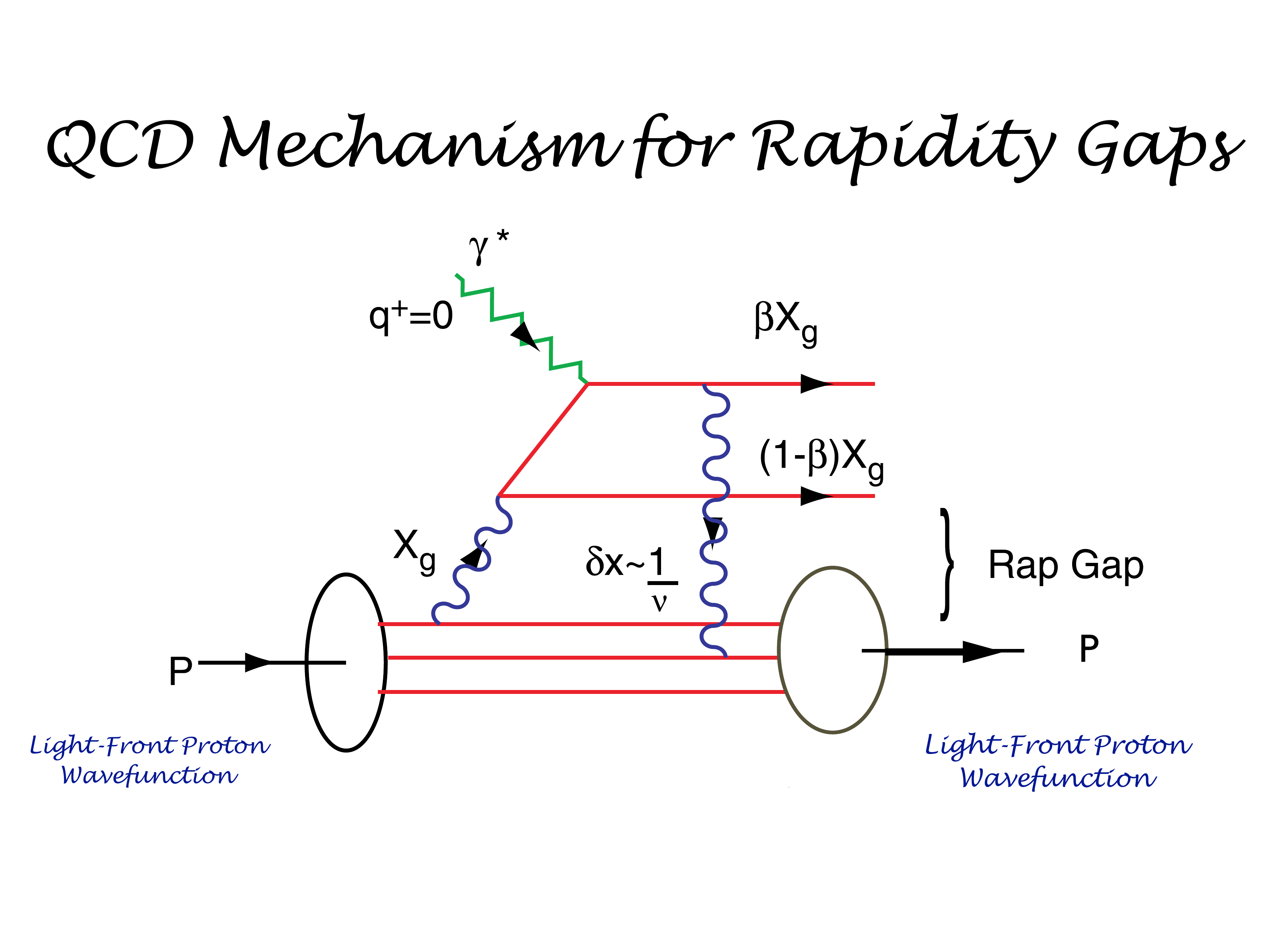}
\end{center}
  \caption{QCD mechanism for diffractive deep inelastic scattering.}
\label{figNew7}  
\end{figure}

A new understanding of the role of final-state interactions in deep inelastic scattering has thus emerged. The multiple scattering of the struck
parton via instantaneous interactions in the target generates dominantly imaginary diffractive amplitudes, giving rise to an effective ``hard
pomeron" exchange.  The presence of a rapidity gap between the target and diffractive system requires that the target remnant emerges in a
color-singlet state; this is made possible in any gauge by the soft rescattering, as illustrated in fig.\ref{figNew7}. The resulting diffractive contributions leave the target
intact  and do not resolve its quark structure; thus there are contributions to the DIS structure functions which cannot be interpreted as
parton probabilities~\cite{Brodsky:2002ue}; the leading-twist contribution to DIS  from rescattering of a quark in the target is a coherent
effect which is not included in the light-front wave functions computed in isolation. One can augment the light-front wave functions with a
gauge link corresponding to an external field created by the virtual photon $q \bar q$ pair current~\cite{Belitsky:2002sm,Collins:2004nx}.  Such
a gauge link is process dependent~\cite{Collins:2002kn}, so the resulting augmented LFWFs are not universal~\cite{Brodsky:2002ue,Belitsky:2002sm,Collins:2003fm}.   We also note that the shadowing of nuclear structure functions is due to the destructive
interference between multi-nucleon amplitudes involving diffractive DIS and on-shell intermediate states with a complex phase. In contrast, the
wave function of a stable target is strictly real since it does not have on-energy-shell intermediate state configurations.  The physics of
rescattering and shadowing is thus not included in the nuclear light-front wave functions, and a probabilistic interpretation of the nuclear DIS
cross section is precluded.

\begin{figure}[htb]
\centering
\includegraphics[width=15.0cm]{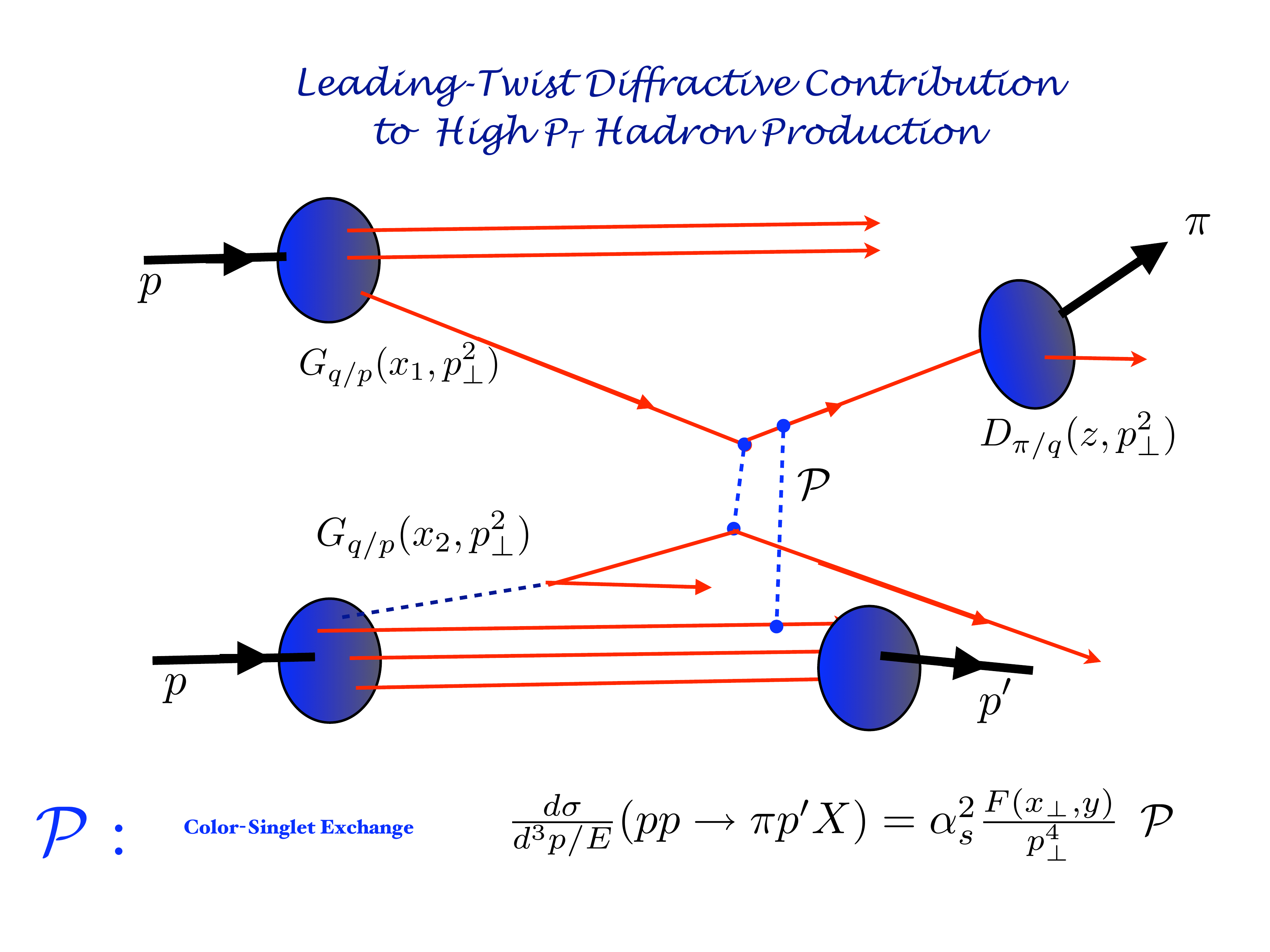}
\caption{Example of a leading-twist diffractive high $p_T$ reactions at the LHC.} \label{figNew18C}
\end{figure}

Rikard Enberg, Paul Hoyer, Gunnar Ingelman and I~\cite{Brodsky:2004hi} have shown that the quark structure function of the effective hard
pomeron has the same form as the quark contribution of the gluon structure function. The hard pomeron is not an intrinsic part of the proton;
rather it must be considered as a dynamical effect of the lepton-proton interaction. Our QCD-based picture also applies to diffraction in
hadron-initiated processes. The rescattering is different in virtual photon- and hadron-induced processes due to the different color
environment, which accounts for the  observed non-universality of diffractive parton distributions. This framework also provides a theoretical
basis for the phenomenologically successful Soft Color Interaction (SCI) model~\cite{Edin:1995gi} which includes rescattering effects and thus
generates a variety of final states with rapidity gaps.
The analogous mechanism will produce leading-twist diffractive high $p_T$ reactions at the LHC, such as 
$ p p \to \pi p X p^\prime.$  See fig.\ref{figNew18C}.

\section{Diffraction Dissociation as a Tool to Resolve Hadron
Substructure and Test Color Transparency}

Diffractive multi-jet production in heavy nuclei provides a novel way to resolve the shape of light-front Fock state wave functions and test
color transparency~\cite{Brodsky:1988xz}.  For example, consider the reaction~\cite{Bertsch:1981py,Frankfurt:1999tq}.   $\pi A \rightarrow {\rm
Jet}_1 + {\rm Jet}_2 + A^\prime$ at high energy where the nucleus $A^\prime$ is left intact in its ground state. The transverse momenta of the
jets balance so that $ \vec k_{\perp i} + \vec k_{\perp 2} = \vec q_\perp < {R^{-1}}_A \ $.  Because of color transparency, the valence wave
function of the pion with small impact separation will penetrate the nucleus with minimal interactions, diffracting into jet
pairs~\cite{Bertsch:1981py}.  The $x_1=x$, $x_2=1-x$ dependence of the dijet distributions will thus reflect the shape of the pion valence
light-cone wave function in $x$; similarly, the $\vec k_{\perp 1}- \vec k_{\perp 2}$ relative transverse momenta of the jets gives key
information on the underlying shape of the valence pion
wavefunction~\cite{Frankfurt:1999tq,Nikolaev:2000sh}. 
The measured
 fractional momentum distribution of the jets with high transverse momentum reflects the shape of the second derivative $\partial^2 \Psi_\pi(x,\vec k_\perp)/ \partial^2 \vec k_\perp $  of the incident pion's LFWF;  the $x$-dependence, determined from the longitudinal momentum fractions of the jets, is found to be approximately consistent with the shape of the pion
asymptotic distribution amplitude $\phi^{\rm asympt}_\pi (x) = \sqrt 3 f_\pi x(1-x)$~\cite{Aitala:2000hb} for high jet $k_T$;  however, there are hints from
the data that the $x$-distribution is considerably broader at lower transverse momentum, consistent with the AdS/CFT expectations discussed below.

Color transparency, as evidenced by the Fermilab E791 measurement of diffractive dijet production, implies that a pion can interact coherently
throughout a nucleus with minimal absorption, in dramatic contrast to traditional Glauber theory based on a fixed $\sigma_{\pi n}$ cross
section.  
Color Transparency~\cite{Brodsky:1988xz}, a key
feature of the gauge theoretic description of hadron interactions, has now been experimentally established at FermiLab~\cite{Ashery:2006zw}
using diffractive dijet production $\pi A \to {\rm jet} {\rm jet} A$. The diffractive nuclear amplitude extrapolated to $t = 0$ should be linear in nuclear
number $A$ if color transparency is correct.  The integrated diffractive rate is predicted to scale as $A^2/R^2_A \sim A^{4/3}$, far different than conventional nuclear theory.  This is in fact what
has been observed by the E791 collaboration at FermiLab for 500 GeV incident pions on nuclear targets~\cite{Aitala:2000hc}.  
An analogous measurement at the LHC $ p A \to Jet Jet Jet A$ at the LHC could be used to
measure the fundamental valence wavefunction of the proton~\cite{Frankfurt:2002pu}.
Color transparency has also been observed in diffractive
electroproduction of $\rho$ mesons~\cite{Borisov:2002rd} and in quasi-elastic $p A \to p p (A-1)$ scattering~\cite{Aclander:2004zm} where only
the small size fluctuations of the hadron wavefunction enters the hard exclusive scattering amplitude.

\section{ Single-Spin Asymmetries from Rescattering}

\begin{figure}[htb]
\centering
\includegraphics[width=15.0cm]{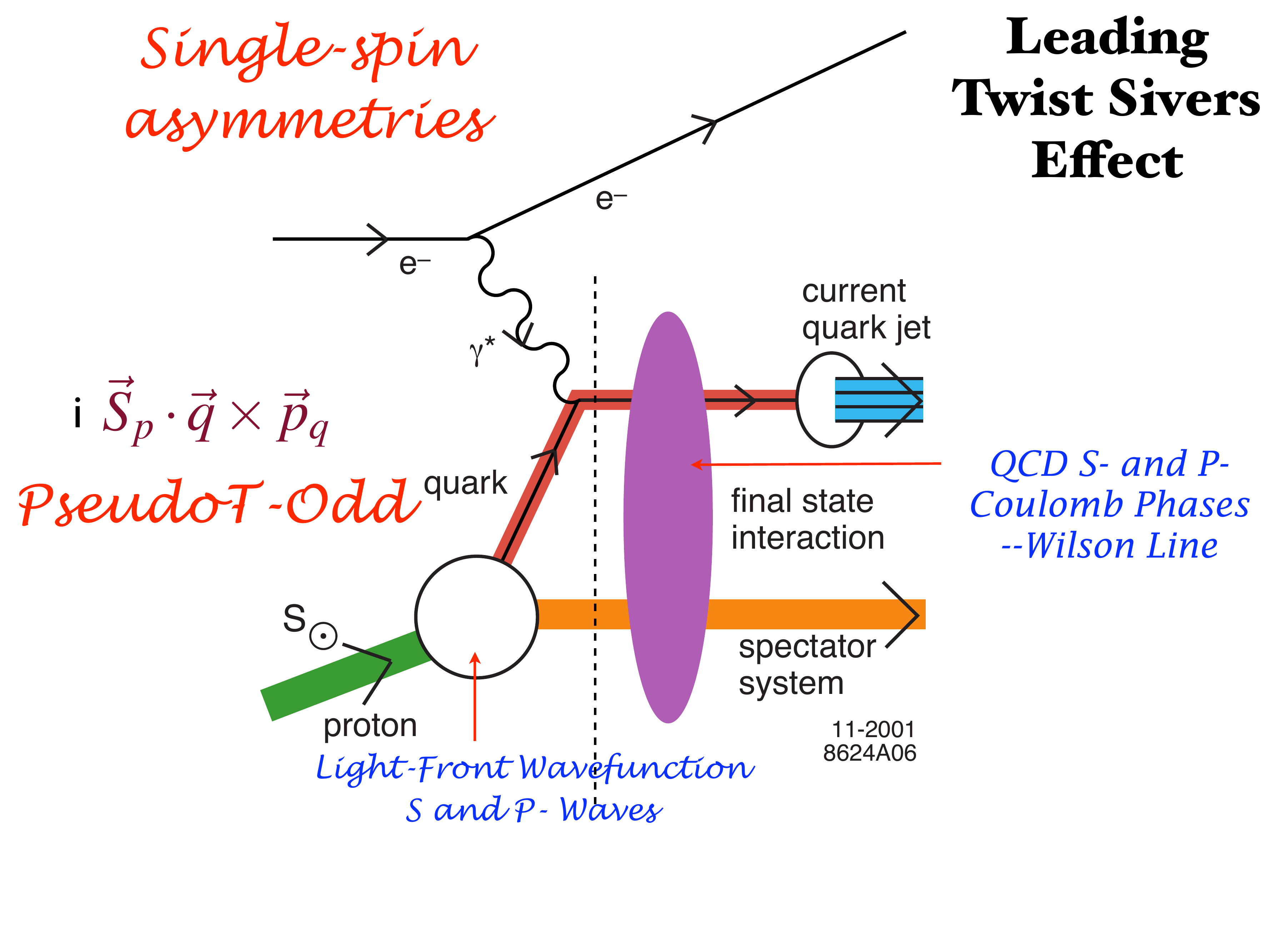}
\caption{Final-state interactions in QCD and the physics of the leading-twist Sivers single-spin asymmetry in semi-inclusive deep inelastic
lepton-proton scattering.} \label{figNew8}
\end{figure}

Among the most interesting polarization effects are single-spin azimuthal asymmetries  in semi-inclusive deep inelastic scattering, representing
the correlation of the spin of the proton target and the virtual photon to hadron production plane: $\vec S_p \cdot \vec q \times \vec p_H$.
Such asymmetries are time-reversal odd, but they can arise in QCD through phase differences in different spin amplitudes. In fact, final-state
interactions from gluon exchange between the outgoing quarks and the target spectator system lead to single-spin asymmetries in semi-inclusive
deep inelastic lepton-proton scattering  which  are not power-law suppressed at large photon virtuality $Q^2$ at fixed
$x_{bj}$~\cite{Brodsky:2002cx}.  In contrast to the SSAs arising from transversity and the Collins fragmentation function, the fragmentation of
the quark into hadrons is not necessary; one predicts a correlation with the production plane of the quark jet itself. Physically, the
final-state interaction phase arises as the infrared-finite difference of QCD Coulomb phases for hadron wave functions with differing orbital
angular momentum.  See fig.\ref{figNew8}. The same proton matrix element which determines the spin-orbit correlation $\vec S \cdot \vec L$ also
produces the anomalous magnetic moment of the proton, the Pauli form factor, and the generalized parton distribution $E$ which is measured in
deeply virtual Compton scattering. Thus the contribution of each quark current to the SSA is proportional to the contribution $\kappa_{q/p}$ of
that quark to the proton target's anomalous magnetic moment $\kappa_p = \sum_q e_q \kappa_{q/p}$ ~\cite{Brodsky:2002cx,Burkardt:2004vm}.  As shown
by Gardner and myself~\cite{Brodsky:2006ha}, one can also use the Sivers effect to study the orbital angular momentum of  gluons by
tagging a gluon jet in semi-inclusive DIS. In this case, the final-state interactions are enhanced by the large color charge of the gluons.

The
HERMES collaboration has measured the single spin asymmetry in pion electroproduction using transverse target polarization~\cite{Airapetian:2004tw}. The
Sivers and Collins effects can be separated using planar correlations; both contributions are observed to contribute, with values not in
disagreement with theory expectations ~\cite{Airapetian:2004tw,Avakian:2004qt}. A related analysis also predicts that the initial-state
interactions from gluon exchange between the incoming quark and the target spectator system lead to leading-twist single-spin asymmetries in the
Drell-Yan process $H_1 H_2^\updownarrow \to \ell^+ \ell^- X$ ~\cite{Collins:2002kn,Brodsky:2002rv}.  The SSA in the Drell-Yan process is the
same as that obtained in SIDIS, with the appropriate identification of variables, but with the opposite sign. There is no Sivers effect in
charged-current reactions since the $W$ only couples to left-handed quarks~\cite{Brodsky:2002pr}.

If both the quark and antiquark in the initial state of the Drell-Yan subprocess $q \bar q \to  \mu^+ \mu^-$ interact with the spectators of the
other incident hadron, one finds a breakdown of the Lam-Tung relation, which was formerly believed to be a general prediction of leading-twist
QCD. These double initial-state interactions also lead to a $\cos 2 \phi$ planar correlation in unpolarized Drell-Yan
reactions~\cite{Boer:2002ju}. More generally one must consider subprocesses involving initial-state gluons such as $n g q \bar q \to \ell \bar
\ell$  in addition to subprocesses with extra final-state gluons.

The final-state interaction effects can also be identified with the gauge link which is present in the gauge-invariant definition of parton
distributions~\cite{Collins:2004nx}.  Even when the light-cone gauge is chosen, a transverse gauge link is required.  Thus in any gauge the
parton amplitudes need to be augmented by an additional eikonal factor incorporating the final-state interaction and its
phase~\cite{Ji:2002aa,Belitsky:2002sm}. The net effect is that it is possible to define transverse momentum dependent parton distribution
functions which contain the effect of the QCD final-state interactions.

\begin{figure}[!]
\includegraphics[width=15cm]{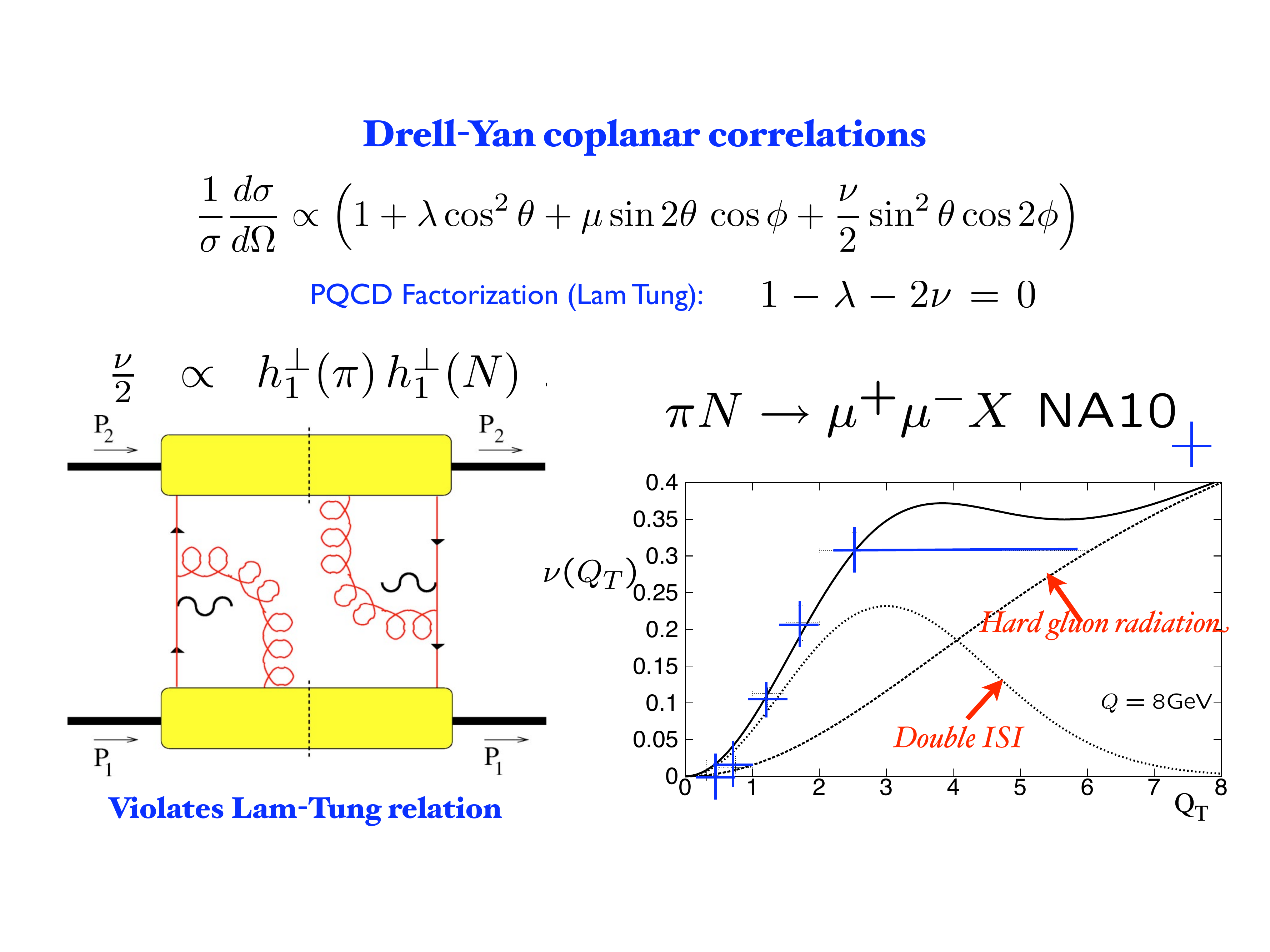}
\caption{Double initial-state interactions break the Lam-Tung relation in Drell-Yan processes~\cite{Boer:2002ju}.}
\label{figNew15A}  
\end{figure}

As  noted by Collins and Qiu~\cite{Collins:2007nk}, the traditional factorization formalism of perturbative QCD for high transverse
momentum hadron production fails in detail because of initial- and final-state gluonic interactions.   If both the
quark and antiquark in the initial state of the Drell-Yan subprocess
$q \bar q \to  \mu^+ \mu^-$ interact with the spectators of the
other incident hadron, one finds a breakdown of the Lam-Tung
relation, which was formerly believed to be a general prediction of
leading-twist QCD. These double initial-state interactions also lead
to a $\cos 2\phi \sin^2 \theta$ planar correlation in unpolarized Drell-Yan
reactions~\cite{Boer:2002ju}.  One can thus account for the large $\cos 2 \phi$ correlation and violation~\cite{Boer:1999mm,Boer:2002ju} of the Lam Tung relation for Drell-Yan processes seen by the NA10 collaboration.   This is illustrated in fig.\ref{figNew15A} .   Zhou et al. ~\cite{Zhou:2009rp} have also shown explicitly that the breakdown of the PQCD factorization is leading twist for $Q^2 \to \infty$ and finite $Q_\perp$ transverse momentum of the lepton pair. 

\begin{figure}[!]
\includegraphics[width=15cm]{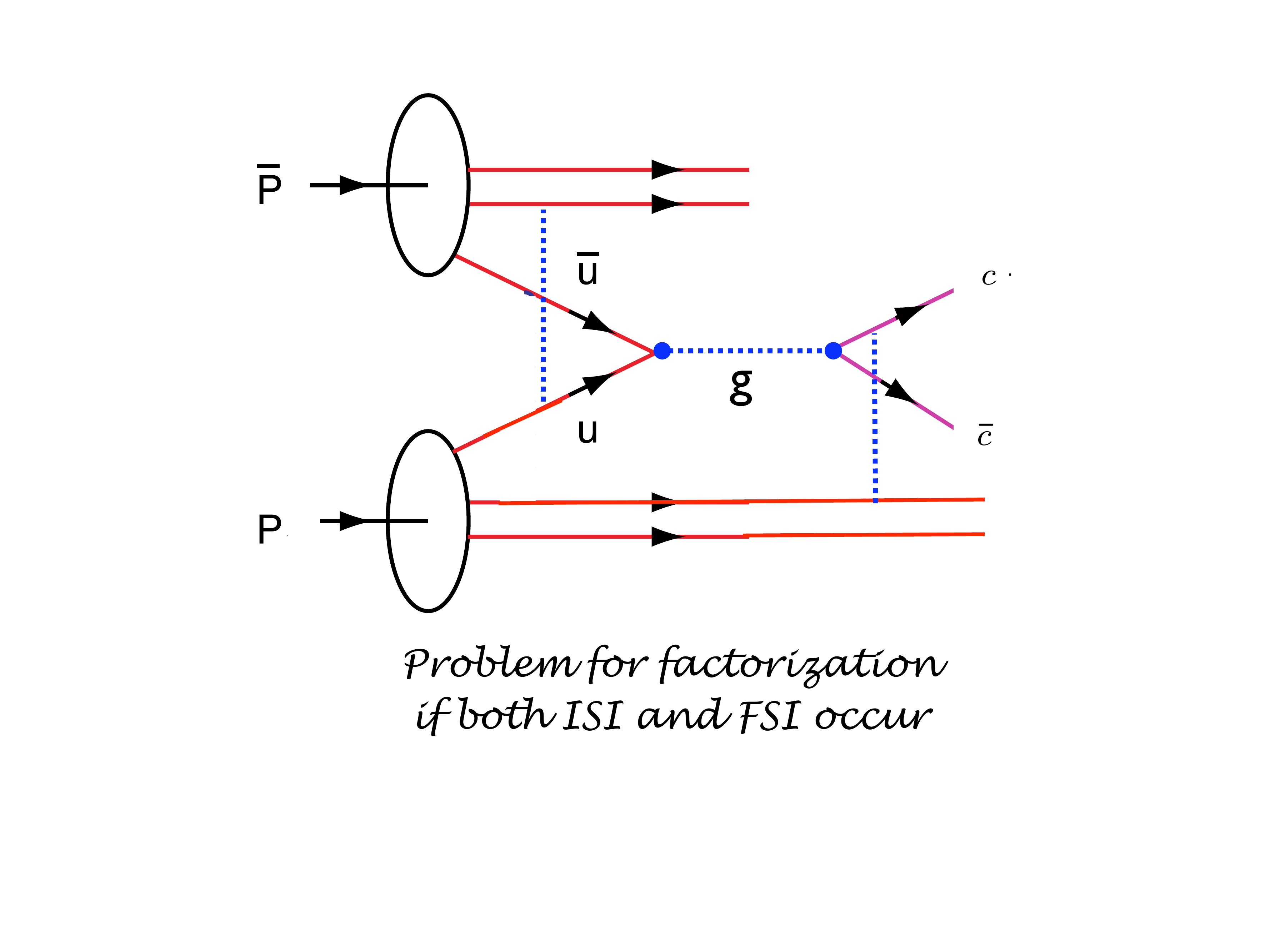}
\caption{Initial-state and  final-state corrections lead to non-factorization of heavy-quark hadroproduction reactions, in analogy to the breakdown  the Lam-Tung Relation in Drell-Yan Processes.  }
\label{figNew15B}  
\end{figure}

The problem of leading twist factorization breakdown is compounded when both initial and final-state interactions occur, as shown in fig.\ref{figNew15B}.
An important signal for factorization breakdown at the LHC  will be the observation of a $\cos 2 \phi$ planar correlation in dijet production.

\section{Leading-Twist Shadowing of Nuclear Structure Functions}

The shadowing of the nuclear structure functions: $R_A(x,Q^2) < 1 $ at small $x < 0.1 $ can be readily understood in terms of the Gribov-Glauber
theory.  Consider a two-step process in the nuclear target rest frame. The incoming $q \bar q$ dipole first interacts diffractively $\gamma^*
N_1 \to (q \bar q) N_1$ on nucleon $N_1$ leaving it intact.  This is the leading-twist diffractive deep inelastic scattering  (DDIS) process
which has been measured at HERA to constitute approximately 10\% of the DIS cross section at high energies.  The $q \bar q$ state then interacts
inelastically on a downstream nucleon $N_2:$ $(q \bar q) N_2 \to X$. The phase of the pomeron-dominated DDIS amplitude is close to imaginary,
and the Glauber cut provides another phase $i$, so that the two-step process has opposite  phase and  destructively interferes with the one-step
DIS process $\gamma* N_2 \to X$ where $N_1$ acts as an unscattered spectator. The one-step and-two-step amplitudes can coherently interfere as
long as the momentum transfer to the nucleon $N_1$ is sufficiently small that it remains in the nuclear target;  {\em i.e.}, the Ioffe
length~\cite{Ioffe:1969kf} $L_I = { 2 M \nu/ Q^2} $ is large compared to the inter-nucleon separation. In effect, the flux reaching the interior
nucleons is diminished, thus reducing the number of effective nucleons and $R_A(x,Q^2) < 1.$

As noted above, the Bjorken-scaling diffractive contribution to DIS arises from the rescattering of the struck quark after it is struck  (in the
parton model frame $q^+ \le 0$), an effect induced by the Wilson line connecting the currents. Thus one cannot attribute DDIS to the physics of
the target nucleon computed in isolation~\cite{Brodsky:2002ue}.  Similarly, since shadowing and antishadowing arise from the physics of
diffraction, we cannot attribute these phenomena to the structure of the nucleus itself: shadowing and antishadowing arise because of the
$\gamma^* A$ collision and the history of the $q \bar q$ dipole as it propagates through the nucleus.

\section{Non-Universal Antishadowing}

One of the novel features of QCD involving nuclei is the {\it antishadowing} of the nuclear structure functions as observed in deep
inelastic lepton-nucleus scattering. Empirically, one finds $R_A(x,Q^2) \equiv  \left(F_{2A}(x,Q^2)/ (A/2) F_{d}(x,Q^2)\right)
> 1 $ in the domain $0.1 < x < 0.2$;  {\em i.e.}, the measured nuclear structure function (referenced to the deuteron) is larger than the
scattering on a set of $A$ independent nucleons.

Ivan Schmidt, Jian-Jun Yang, and I~\cite{Brodsky:2004qa} have extended the analysis of nuclear shadowing  to the shadowing and antishadowing of the
electroweak structure functions.  We note that there are leading-twist diffractive contributions $\gamma^* N_1 \to (q \bar q) N_1$  arising from Reggeon exchanges in the
$t$-channel~\cite{Brodsky:1989qz}.  For example, isospin--non-singlet $C=+$ Reggeons contribute to the difference of proton and neutron
structure functions, giving the characteristic Kuti-Weisskopf $F_{2p} - F_{2n} \sim x^{1-\alpha_R(0)} \sim x^{0.5}$ behavior at small $x$. The
$x$ dependence of the structure functions reflects the Regge behavior $\nu^{\alpha_R(0)} $ of the virtual Compton amplitude at fixed $Q^2$ and
$t=0.$ The phase of the diffractive amplitude is determined by analyticity and crossing to be proportional to $-1+ i$ for $\alpha_R=0.5,$ which
together with the phase from the Glauber cut, leads to {\it constructive} interference of the diffractive and nondiffractive multi-step nuclear
amplitudes.  The nuclear structure function is predicted to be enhanced precisely in the domain $0.1 < x <0.2$ where
antishadowing is empirically observed.  The strength of the Reggeon amplitudes is fixed by the fits to the nucleon structure functions, so there
is little model dependence.
Since quarks of different flavors  will couple to different Reggeons; this leads to the remarkable prediction that
nuclear antishadowing is not universal; it depends on the quantum numbers of the struck quark. This picture implies substantially different
antishadowing for charged and neutral current reactions, thus affecting the extraction of the weak-mixing angle $\theta_W$.  The ratio of nuclear to nucleon structure functions $R_{A/N}(x,Q) = {F_{2A}(x,Q)\over A F_{2N}(x,Q)}$ is thus process independent.   We have also identified
contributions to the nuclear multi-step reactions which arise from odderon exchange and hidden color degrees of freedom in the nuclear
wavefunction.

\begin{figure}[!]
 \begin{center}
\includegraphics[width=15.0cm]{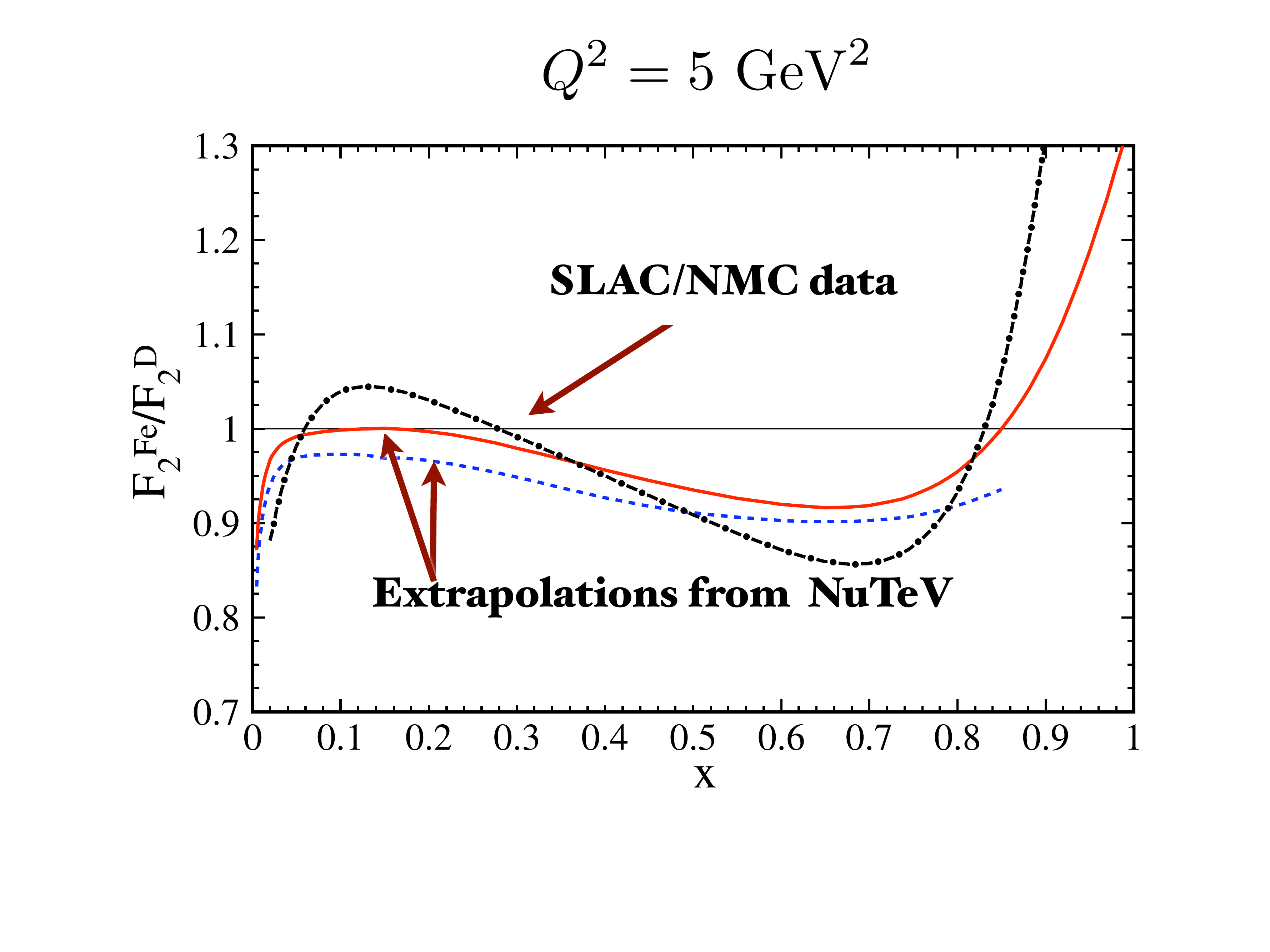}
\end{center}
  \caption{Comparison of the Nuclear Modification  of Charged vs. Neutral Current Deep Inelastic Structure Functions.  From Olness et al. }
\label{figNew9}  
\end{figure}

Schienbein et al. ~\cite{Schienbein:2008ay} have recently given a comprehensive analysis of charged current deep inelastic neutrino-iron scattering, finding significant differences with the nuclear corrections for electron-iron scattering.  See fig.\ref{figNew9}.  The measured nuclear effect measured in the NuTeV deep inelastic scattering charged current experiment  is distinctly different from the nuclear modification measured at SLAC and NMC in deep inelastic scattering electron and muon scattering.     This implies that part of
of the anomalous NuTeV result~\cite{Zeller:2001hh} for $\theta_W$ could be due to the non-universality of nuclear antishadowing for charged and
neutral currents.

A new understanding of nuclear shadowing and antishadowing has  emerged based on multi-step coherent reactions involving
leading twist diffractive reactions~\cite{Brodsky:1989qz,Brodsky:2004qa}. The nuclear shadowing of structure functions is a consequence of
the lepton-nucleus collision; it is not an intrinsic property of the nuclear wavefunction. The same analysis shows that antishadowing is {\it
not universal}, but it depends in detail on the flavor of the quark or antiquark constituent~\cite{Brodsky:2004qa}.

Detailed measurements of the nuclear dependence of individual quark structure functions are thus needed to establish the
distinctive phenomenology of shadowing and antishadowing and to make the NuTeV results definitive.  A comparison of the nuclear modification in neutrino versus anti-neutrino interactions is clearly important.  There are other ways in which this new view of antishadowing can be tested;  for example, antishadowing can also depend on the target and beam
polarization.

\begin{figure}[!]
\includegraphics[width=15cm]{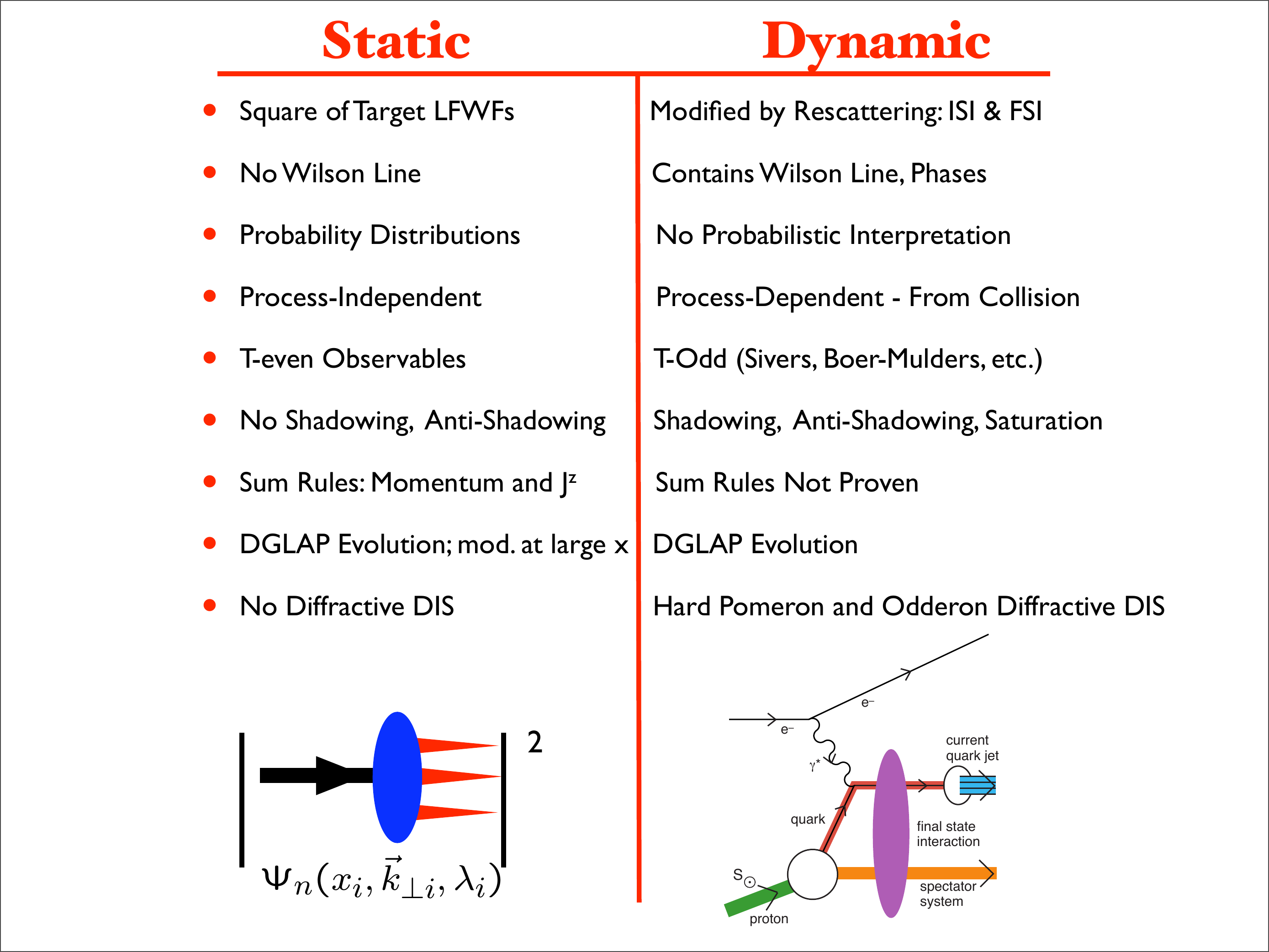}
\caption{Static versus dynamic structure functions}
\label{figNew17}  
\end{figure}

\section{Dynamic versus Static Hadronic Structure Functions}
The nontrivial effects from rescattering and diffraction highlight the need for a fundamental understanding the dynamics of hadrons in QCD at the amplitude
level. This is essential for understanding phenomena such as the quantum mechanics of hadron formation, the remarkable
effects of initial and final interactions, the origins of diffractive phenomena and single-spin asymmetries, and manifestations of higher-twist
semi-exclusive hadron subprocesses. A central tool in these analyses is the light-front wavefunctions of hadrons, the frame-independent
eigensolutions of the Heisenberg equation for QCD ~  $H^{LF}|\Psi> = M^2 |\Psi>$ quantized at fixed light-front. Given the light-front
wavefunctions $\psi_{n/H}(x_i, \vec k_{\perp i}, \lambda_i )$, one can compute a large range of exclusive and inclusive hadron observables. For
example, the valence, sea-quark and gluon distributions are defined from the squares of the LFWFS summed over all Fock states $n$. Form factors,
exclusive weak transition amplitudes~\cite{Brodsky:1998hn} such as $B\to \ell \nu \pi,$ and the generalized parton
distributions~\cite{Brodsky:2000xy} measured in deeply virtual Compton scattering are (assuming the ``handbag" approximation) overlaps of the
initial and final LFWFS with $n =n^\prime$ and $n =n^\prime+2$.

It is thus important to distinguish ``static" structure functions which are computed directly from the light-front wavefunctions of  a target hadron from the nonuniversal ``dynamic" empirical structure functions which take into account rescattering of the struck quark in deep inelastic lepton scattering. 
See  fig.\ref{figNew17}.
The real wavefunctions underlying static structure functions cannot describe diffractive deep inelastic scattering nor  single-spin asymmetries, since such phenomena involve the complex phase structure of the $\gamma^* p $ amplitude.   
One can augment the light-front wavefunctions with a gauge link corresponding to an external field
created by the virtual photon $q \bar q$ pair
current~\cite{Belitsky:2002sm,Collins:2004nx}, but such a gauge link is
process dependent~\cite{Collins:2002kn}, so the resulting augmented
wavefunctions are not universal.
\cite{Brodsky:2002ue,Belitsky:2002sm,Collins:2003fm}.  The physics of rescattering and nuclear shadowing is not
included in the nuclear light-front wavefunctions, and a
probabilistic interpretation of the nuclear DIS cross section is
precluded.

\section{Novel Intrinsic Heavy Quark Phenomena}

Intrinsic heavy quark distributions are a rigorous feature of QCD, arising from diagrams in which two or more gluons couple the valence quarks to the heavy quarks.
The probability for Fock states of a light hadron to have an extra heavy quark pair decreases as $1/m^2_Q$ in non-Abelian
gauge theory~\cite{Franz:2000ee,Brodsky:1984nx}.  The relevant matrix element is the cube of the QCD field strength $G^3_{\mu \nu},$  in
contrast to QED where the relevant operator is $F^4_{\mu \nu}$ and the probability of intrinsic heavy leptons in an atomic
state is suppressed as $1/m^4_\ell.$  The maximum probability occurs at $x_i = { m^i_\perp /\sum^n_{j = 1}
m^j_\perp}$ where $m_{\perp i}= \sqrt{k^2_{\perp i} + m^2_i}.$; {\em i.e.}, when the constituents have minimal invariant mass and equal rapidity. Thus the heaviest constituents have the highest
momentum fractions and the highest $x_i$.  Intrinsic charm thus predicts that the charm structure function has support at large $x_{bj}$ in
excess of DGLAP extrapolations~\cite{Brodsky:1980pb}; this is in agreement with the EMC measurements~\cite{Harris:1995jx}. Intrinsic charm can
also explain the $J/\psi \to \rho \pi$ puzzle~\cite{Brodsky:1997fj}. It also affects the extraction of suppressed CKM matrix elements in $B$
decays~\cite{Brodsky:2001yt}.
The dissociation of the intrinsic charm $|uud c \bar c>$ Fock state of the proton can produce a leading heavy quarkonium state at
high $x_F = x_c + x_{\bar c}~$ in $p N \to J/\psi X A^\prime$ since the $c$ and $\bar c$ can readily coalesce into the charmonium state.  Since
the constituents of a given intrinsic heavy-quark Fock state tend to have the same rapidity, coalescence of multiple partons from the projectile
Fock state into charmed hadrons and mesons is also favored.  For example, one can produce a leading $\Lambda_c$ at high $x_F$ and low $p_T$ from
the coalescence of the $u d c$ constituents of the projectile $|uud c \bar c>$  Fock state.

\begin{figure}[!]
 \begin{center}
\includegraphics[width=15.0cm]{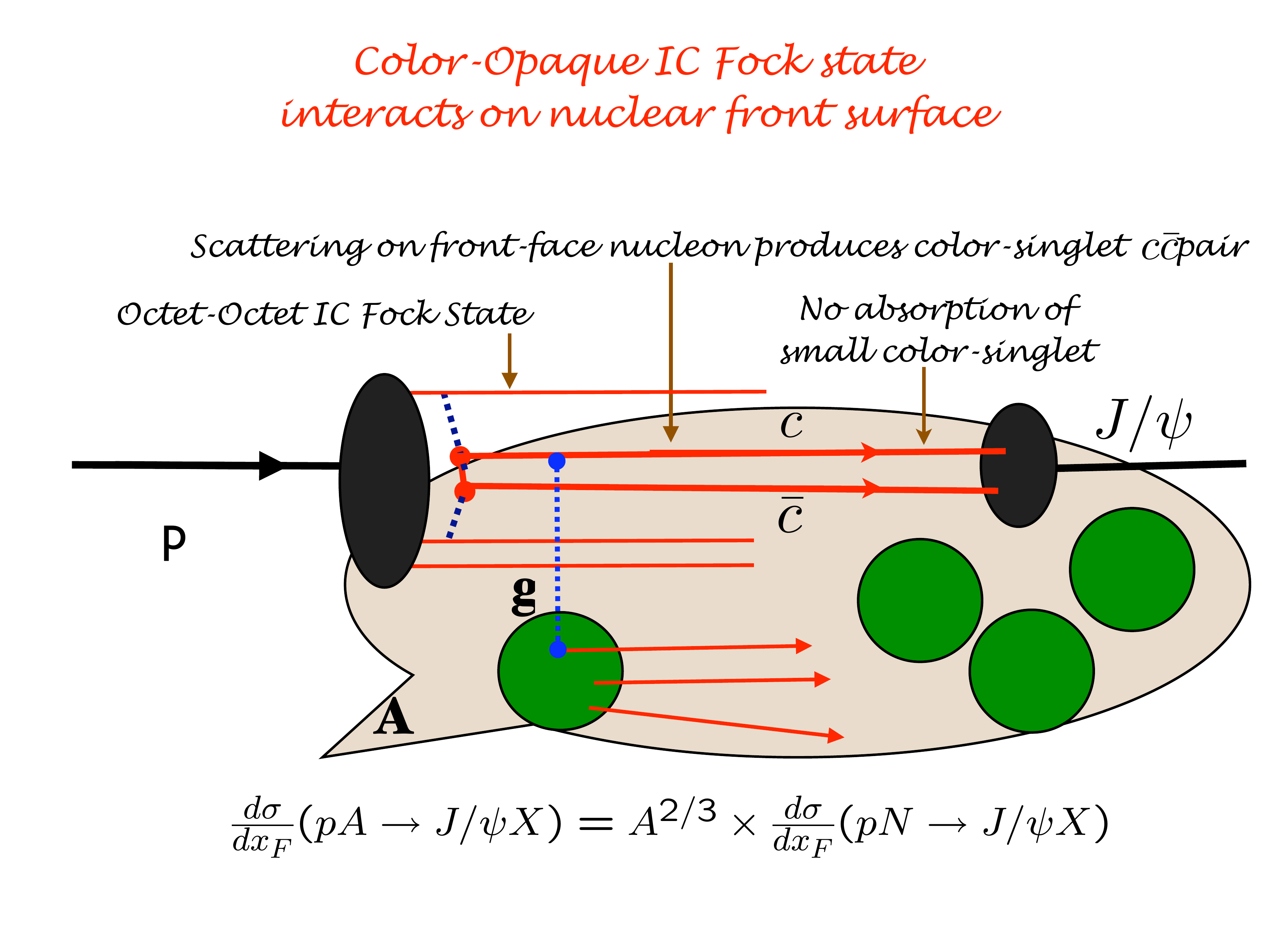}
\end{center}
  \caption{Color-Octet intrinsic charm mechanism for the nuclear dependence of $J/\psi$ production}
\label{figNew10}  
\end{figure}

The operator product analysis of the IC matrix element shows that the IC Fock state has a  dominant color-octet structure: 
$\vert(uud)_{8C} (c \bar c)_{8C}>$. The color octet $c \bar c$ converts to a color singlet by gluon exchange on the front surface of a nuclear target and then coalesces to a $J/\psi$ which interacts weakly through the nuclear volume~\cite{Brodsky:2006wb}.  Thus the rate for the IC component has  $A^{2/3}$ dependence corresponding to the area of the front surface.   This is illustrated in fig \ref{figNew10}. 
This forward contribution is in addition to the $A^1$ contribution derived from the usual perturbative QCD
fusion contribution at small $x_F.$   Because of these two components, the cross section violates perturbative QCD factorization for hard
inclusive reactions~\cite{Hoyer:1990us}.  This is consistent with the two-component cross section for charmonium production observed by
the NA3 collaboration at CERN~\cite{Badier:1981ci} and more recent experiments~\cite{Leitch:1999ea}. The diffractive dissociation of the
intrinsic charm Fock state leads to leading charm hadron production and fast charmonium production in agreement with
measurements~\cite{Anjos:2001jr}.  The hadroproduction cross sections for  double-charm $\Xi_{cc}^+$ baryons at SELEX~\cite{Ocherashvili:2004hi} and the production of $J/\psi$ pairs at NA3 are
 consistent with the diffractive dissociation and coalescence of double IC Fock states~\cite{Vogt:1995tf}. These observations provide
compelling evidence for the diffractive dissociation of complex off-shell Fock states of the projectile and contradict the traditional view that
sea quarks and gluons are always produced perturbatively via DGLAP evolution or gluon splitting.  It is also conceivable that the observations~\cite{Bari:1991ty} of
$\Lambda_b$ at high $x_F$ at the ISR in high energy $p p$  collisions could be due to the dissociation and coalescence of the
``intrinsic bottom" $|uud b \bar b>$ Fock states of the proton.

As emphasized by Lai, Tung, and Pumplin~\cite{Pumplin:2007wg}, there are strong indications that the structure functions used to model charm
and bottom quarks in the proton at large $x_{bj}$ have been underestimated, since they ignore intrinsic heavy quark fluctuations of
hadron wavefunctions. 
Furthermore, the neglect of the intrinsic-heavy quark component  in the proton structure function will lead to an incorrect assessment of the gluon distribution at large $x$ if it is assumed that sea quarks always arise from gluon splitting. It is thus critical for new experiments (HERMES, HERA, COMPASS) to definitively establish the phenomenology of the charm structure function at
large $x_{bj}.$

\begin{figure}[htb]
\centering
\includegraphics[width=15.0cm]{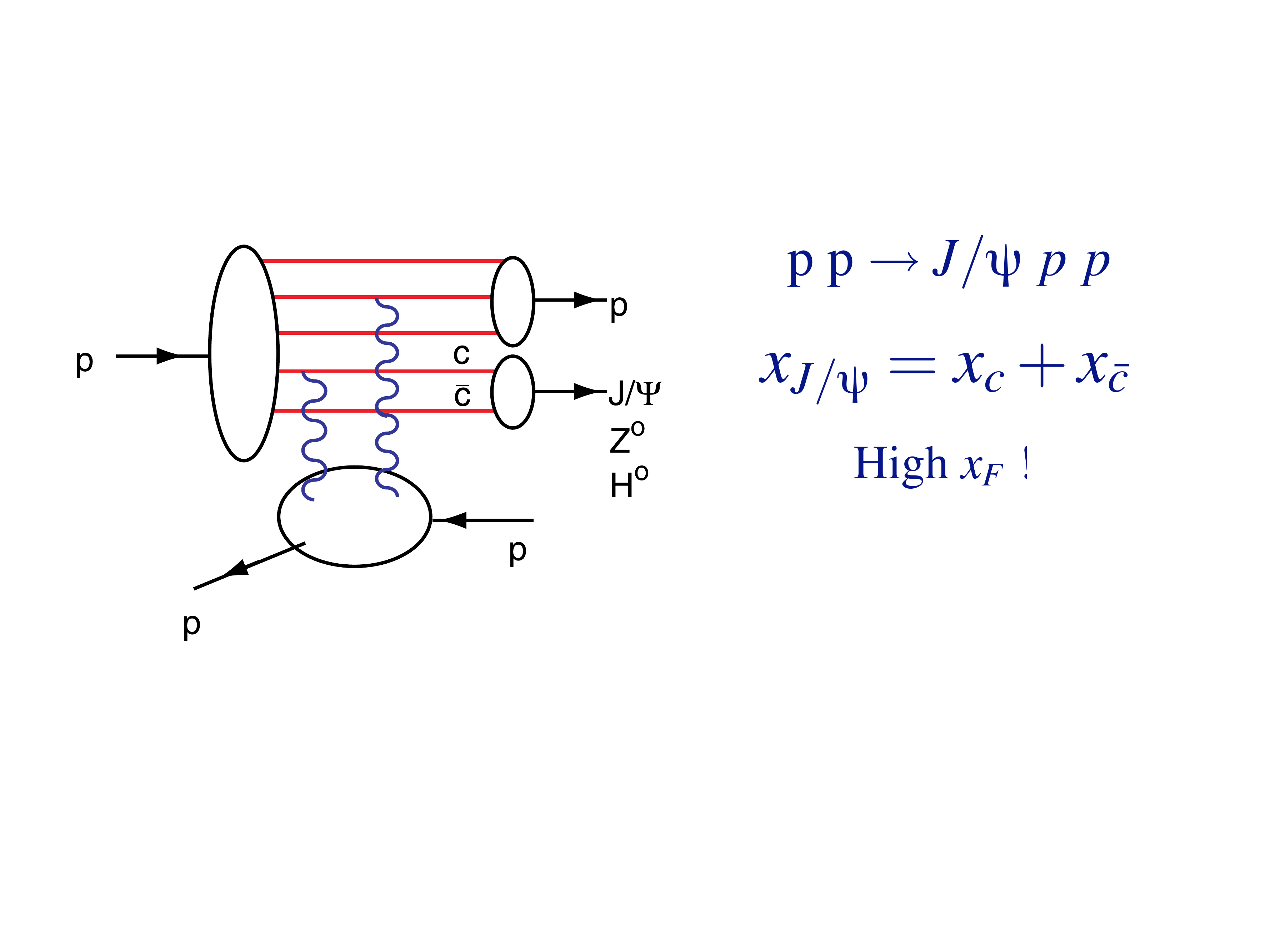}
\caption{Intrinsic heavy quark mechanism for doubly diffractive high $x_F$ Higgs, $Z^0$ and $J/\psi $ production.  } \label{figNew11}
\end{figure}

\begin{figure}[htb]
\centering
\includegraphics[width=15.0cm]{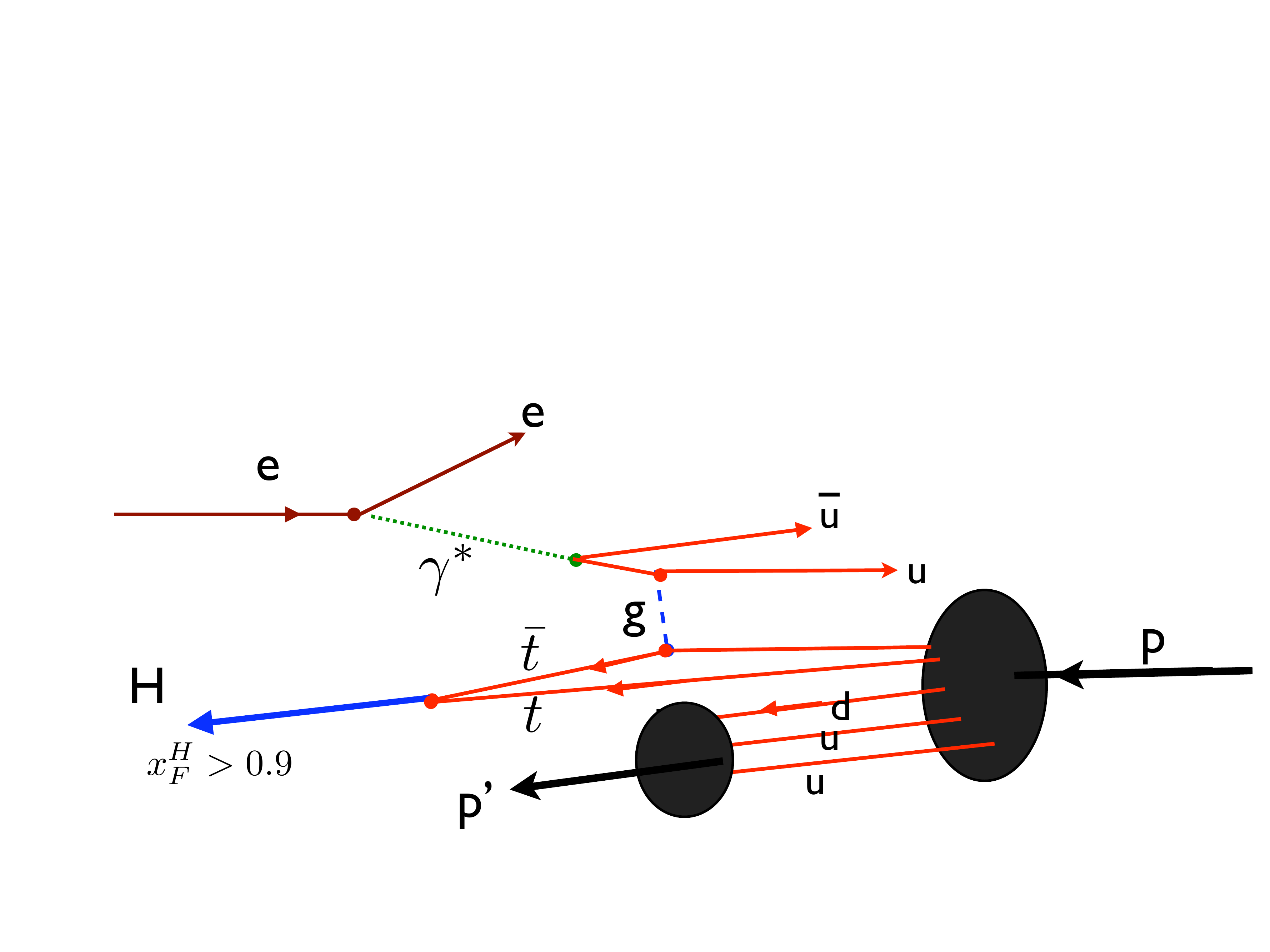}
\caption{Intrinsic heavy quark mechanism for  high $x_F$ Higgs production in electron-proton discussions. } \label{figNew12}
\end{figure}

Intrinsic heavy quarks can also enhance the production probability of Higgs bosons at hadron colliders from processes such as $g c \to H c.$   Recently Kopeliovich, Schmidt, Soffer, and I ~\cite{Brodsky:2006wb} have  proposed a novel mechanism for exclusive diffractive
Higgs production $pp \to p H p $  in which the Higgs boson carries a significant fraction of the projectile proton momentum. The production
mechanism is based on the subprocess $(Q \bar Q) g \to H $ where the $Q \bar Q$ in the $|uud Q \bar Q>$ intrinsic heavy quark Fock state has up
to $80\%$ of the projectile protons momentum. This process, which is illustrated in fig.\ref{figNew11},  will provide a clear experimental signal
for Higgs production due to the small background in this kinematic region. Forward Higgs production from intrinsic heavy quarks at an electron-proton collider 
such as the LHeC is illustrated in fig.\ref{figNew12}.

\section{Hidden Color}
In traditional nuclear physics, the deuteron is a bound state of a proton and a neutron where the binding force arise from the exchange of a
pion and other mesonic states. However, QCD provides a new perspective:~\cite{Brodsky:1976rz,Matveev:1977xt}  six quarks in the fundamental
$3_C$ representation of $SU(3)$ color can combine into five different color-singlet combinations, only one of which corresponds to a proton and
neutron.  In fact, if the deuteron wavefunction is a proton-neutron bound state at large distances, then as their separation becomes smaller,
the QCD evolution resulting from colored gluon exchange introduce four other ``hidden color" states into the deuteron
wavefunction~\cite{Brodsky:1983vf}. The normalization of the deuteron form factor observed at large $Q^2$~\cite{Arnold:1975dd}, as well as the
presence of two mass scales in the scaling behavior of the reduced deuteron form factor~\cite{Brodsky:1976rz}, thus suggest sizable hidden-color
Fock state contributions such as $\ket{(uud)_{8_C} (ddu)_{8_C}}$ with probability  of order $15\%$ in the deuteron
wavefunction~\cite{Farrar:1991qi}.

The hidden color states of the deuteron can be materialized at the hadron level as \\  $\Delta^{++}(uuu)\Delta^{-}(ddd)$ and other novel quantum
fluctuations of the deuteron. These dual hadronic components become more and more important as one probes the deuteron at short distances, such
as in exclusive reactions at large momentum transfer.  For example, the ratio  $${{d \sigma/ dt}(\gamma d \to \Delta^{++}
\Delta^{-})/{d\sigma/dt}(\gamma d\to n p) }$$ should increase dramatically to  a fixed ratio $2::5$ with increasing transverse momentum $p_T.$
Similarly the Coulomb dissociation of the deuteron into various exclusive channels $e d \to e^\prime + p n, p p \pi^-, \Delta \Delta, \cdots$
should have a changing composition as the final-state hadrons are probed at high transverse momentum, reflecting the onset of hidden color
degrees of freedom.

The CLEO collaboration~\cite{Asner:2006pw} has measured the branching ratio for $\Upsilon(nS) \to \ {\rm anti-deuteron} X$. This
reaction should be sensitive to the hidden-color structure of the anti-deuteron wavefunction since the $\Upsilon \to b \bar b \to ggg \to qqqqqq
\bar q \bar q \bar q \bar q \bar q \bar q$ subprocess originates from a system of small compact size and leads to multi-quark states with diverse colors.
It is crucial to have for comparison data on $ \Upsilon \to \bar p \bar n X$ where the anti-nucleons emerge at minimal invariant mass.

\section{Quantum Aspects of Heavy Ion Collisions}

The hadronic system produced in a heavy ion collision has been characterized as a ``holographic quantum liquid"~\cite{Karch:2009zz}
because the empirical value the ratio of shear viscosity to entropy  ${\eta/ S} = {\hbar /4 \pi}$ is not far from the AdS/CFT prediction. The AdS/CFT analysis~\cite{Karch:2009zz} uses a black hole metric in the AdS fifth dimension with a singularity at $z=z_0 = { \hbar  c / T}$ to simulate temperature, thus introducing quantum mechanical scale units characterized by $\hbar.$ 
This  quantum physics result seems counter-intuitive since the multiplicity of hadrons produced in a central Gold-Gold collision can range up to several thousand. 

However, there are a number of aspects of high energy nuclear collisions which are quantum mechanical in nature.  For example, diffractive Processes, shadowing, antishadowing all involve the interference of multistep nuclear amplitudes within a quantum mechanical coherence length. Similarly, the Landau-Pomeronchuk-Migdal effect eliminates radiation between scattering centers because of the destructive interference of multistep processes. Since the energy loss of a quark is finite, this permits a factorized jet fragmentation function.  
Other quantum mechanical effects in nuclei include, color transparency,``hidden color configurations,  $x> 1$ phenomena and other aspects of short-range correlations.

\begin{figure}[!]
 \begin{center}
\includegraphics[width=15.0cm]{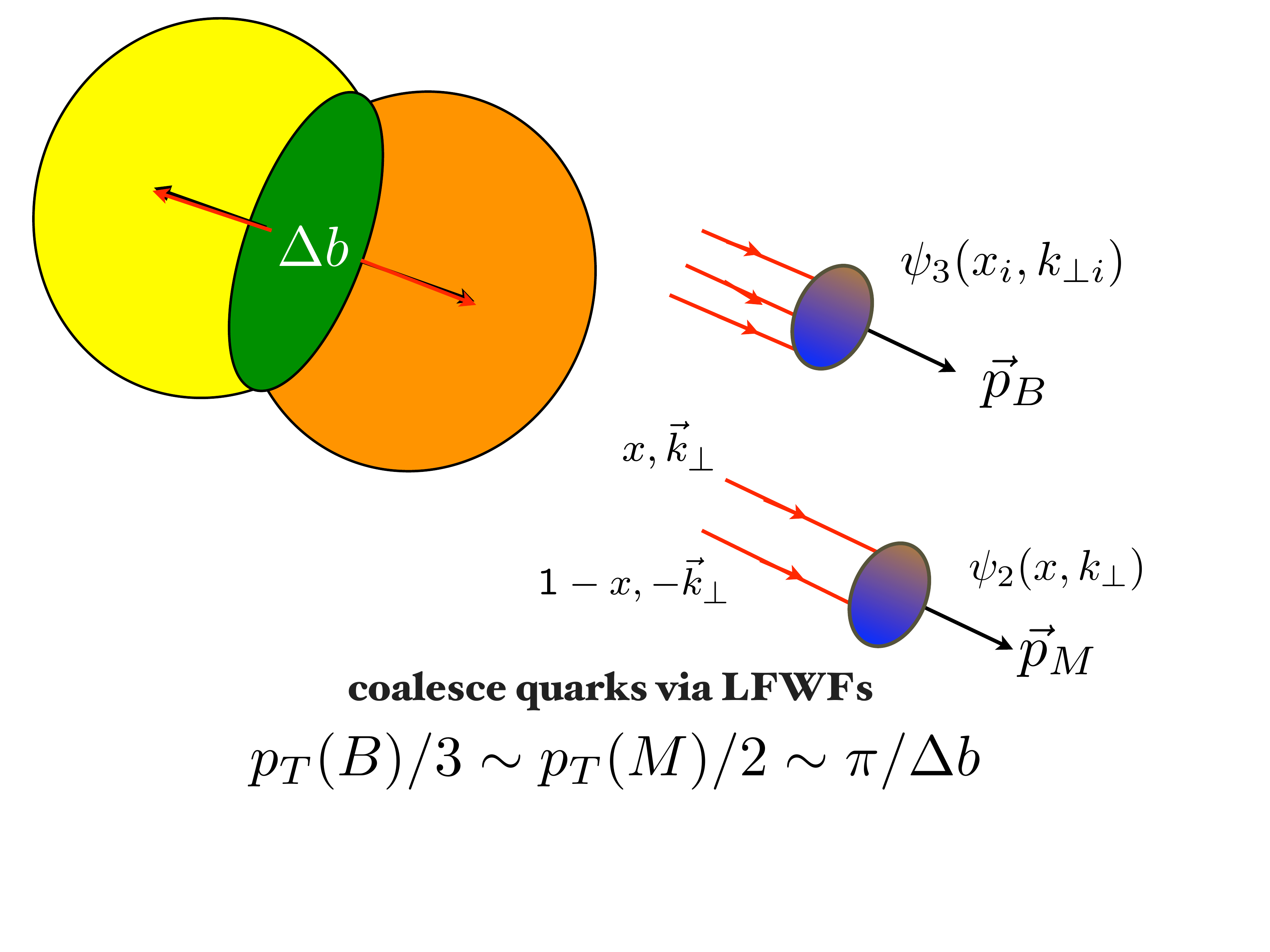}
\end{center}
  \caption{Correlations with Minor Axis of the Overlap Ellipse}
\label{figNew16}  
\end{figure}

If the nucleus-nucleus scattering process is quantum mechanical, then the production of hadrons transverse to the minor axis of the overlap ellipse will have analogy with the quantum mechanical diffractive pattern of light scattering through a slit.    A simplified model is illustrated in fig.\ref{figNew16} .

\begin{figure}[!]
 \begin{center}
\includegraphics[width=15.0cm]{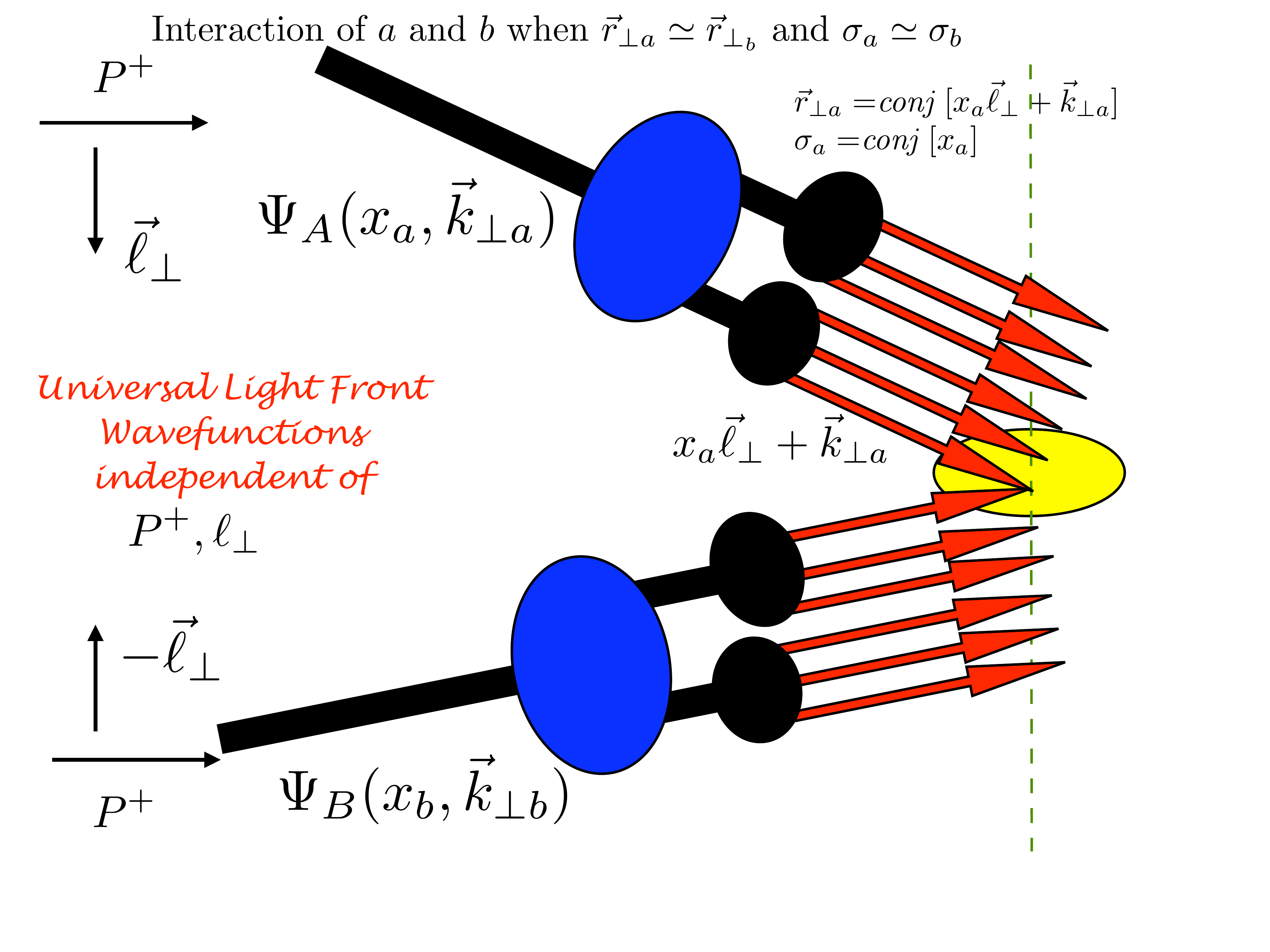}
\end{center}
  \caption{Frame-Independent Description of Nuclear Collisions using Light-Front Wavefunctions}
\label{figNew14}  
\end{figure}

One can use the frame-independent light-front wavefunctions of the incident nuclei to calculate scattering independent of the observer's Lorentz frame. The kinematics used in the light-front analysis is illustrated in fig.\ref{figNew14}.

\begin{figure}[!]
 \begin{center}
\includegraphics[width=15.0cm]{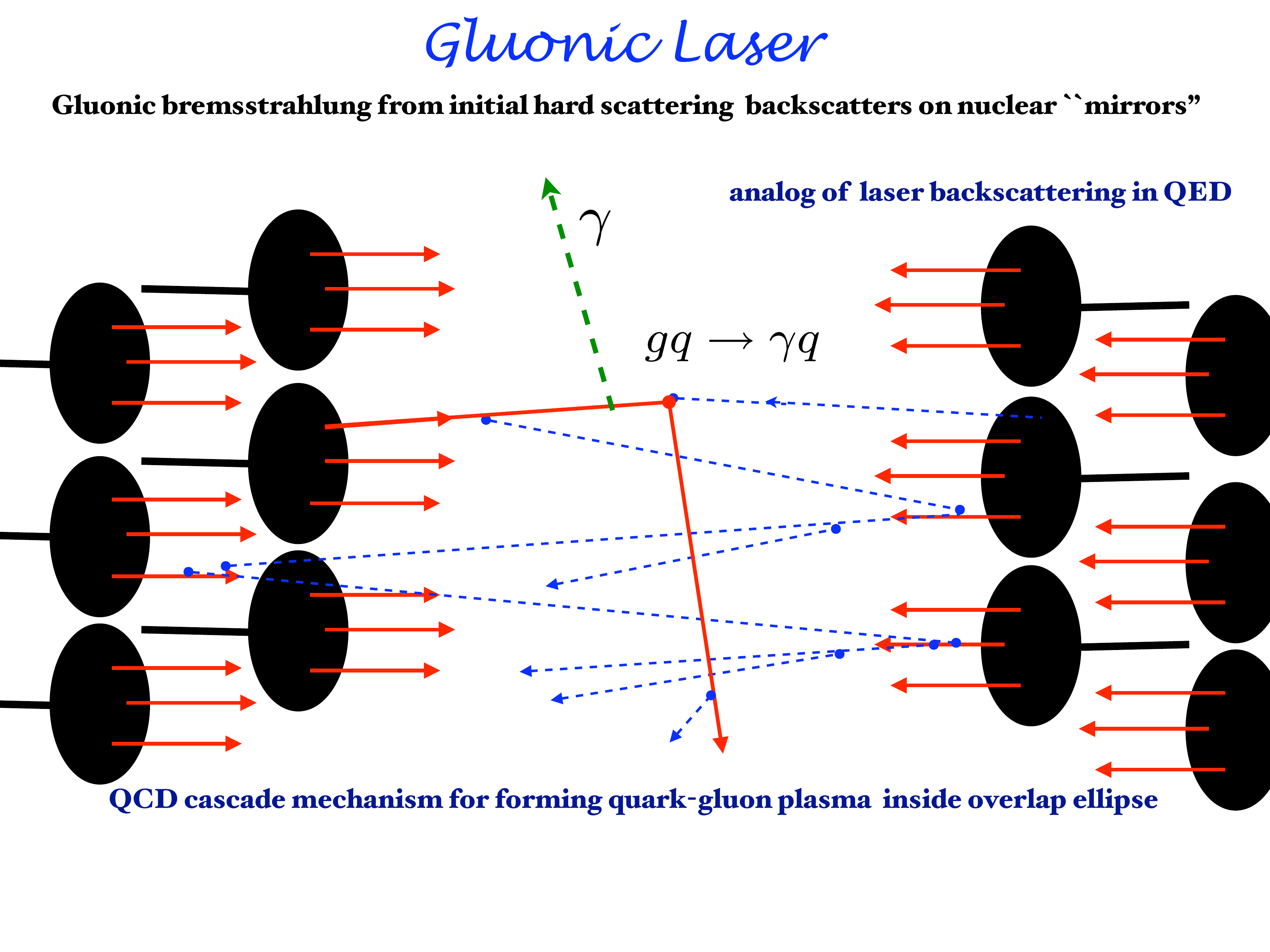}
\end{center}
  \caption{QCD model for early stage thermalization. The reflections only occur at high centrality.}
\label{figNew15}  
\end{figure}

One of the main questions in heavy ion physics is the mechanism which allows thermalization of the interacting quarks and gluons at the early stage of the collisions   A model involving multiple rescattering is illustrated in fig.\ref{figNew15}.  It is  based on the same mechanism which is used to produced a high energy photon beam from laser back-scattering on an electron beam.   Such a model could be used to quantify the conditions needed for setting up an initial quark-gluon plasma system, such as the degree of centrality.

\section{The Ridge in Heavy Ion Collisions}

One of the most interesting phenomena seen at RHIC is the ``ridge" of correlated hadrons which are produced approximately coplanar on the same side of  a high-$p_T$ trigger particle. Remarkably, the ridge extends over a large rapidity interval along the beam direction, so it is not a natural consequence of the jet cone.  The ridge only appears in a high centrality ion-ion collision.    However, as discussed by Barannikova~\cite{Olga} in these proceedings, the ridge does not  appear in dihadron trigger events where the detected high $p_T$ hadrons approximately balance in $p_T.$

\begin{figure}[!]
 \begin{center}
\includegraphics[width=15.0cm]{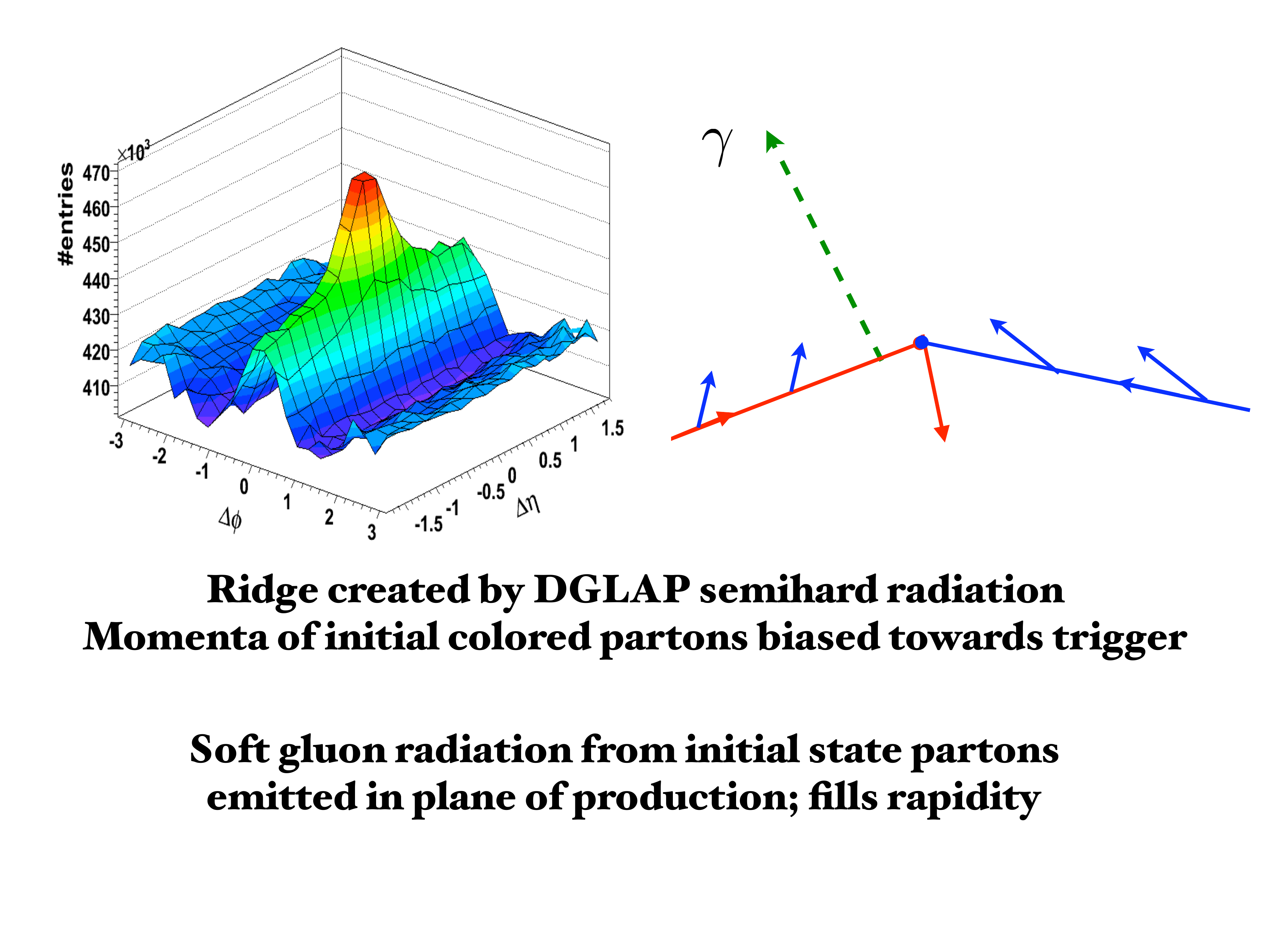}
\end{center}
  \caption{Model for Ridge Production}
\label{figNew13}  
\end{figure}

Dumitru, et al.~\cite{Dumitru:2008wn} have emphasized  that the large rapidity range $\Delta y$ of the ridge requires an early time for its formation: 
$t_o \propto  \exp(-\Delta y)$.  Thus  the ridge could  in fact be a manifestation of the initial-state gluonic radiation associated with the evolution of the incident structure functions.  For definiteness, consider a single hadron event originating from a quark-quark collision: $q_a q_b \to q_c q_d  \to \pi X$.    The initial quarks $q_a$ and $q_b$ each radiate gluons with an approximate $dx/ x$  distribution, where $y = \ln x$ is the rapidity difference between the quark and gluon. This is illustrated in fig.\ref{figNew13}. Thus the DGLAP radiation has a nearly uniform distribution in rapidity.    The gluonic radiation pattern in QCD 
is discussed by Bertsch and Gunion~\cite{Gunion:1981qs}.

DGLAP radiation has a ${\alpha(\ell^2_\perp) /\ell^2_\perp }$ distribution in transverse momentum  relative to the incident quark directions.  The upper limit of the $\vec \ell_\perp$  integration   is  $p_T^2$,  leading to the logarithmic  $\log p^2_T$ DGLAP evolution of the incident structure functions.   In this leading-order analysis there is no correlation between the trigger $p_T$ and the range of the ridge in rapidity.

The momenta of the initial quarks $\vec k_a$ for $q_a$ and $\vec k_b$ for $q_b$ are skewed towards the trigger pion momentum. The semihard gluons radiated by $q_a$ and $q_b$  are thus also skewed toward the trigger.  These gluons can interact with the hadronic matter produced in the central collision. For example, the radiated gluon could coalesce with a gluon in the background medium to produce a gluonium state moving toward the trigger; thus the initial state radiation can imprint the nuclear medium and produce the ridge correlated in $\phi$ with the trigger, and extending over rapidity.  

The trigger bias in parton $k_T$ vanishes in the case of dihadron triggers which balance in $p_T$, thus explaining the RHIC observation~\cite{Olga}. On the other hand, a ridge extending over rapidity should be formed in correlation with a direct photon, although the strength of the ridge correlation could be reduced relative to a hadron trigger.

\section{AdS/QCD as a First Approximant to Nonperturbative QCD}

One of the most interesting new developments in hadron physics has been the application of the AdS/CFT correspondence~\cite{Maldacena:1997re} to
nonperturbative QCD 
problems
~\cite{Polchinski:2001tt,Janik:1999zk,Erlich:2005qh,Karch:2006pv,deTeramond:2005su,deTeramond:2008ht}.   
The application of AdS space and conformal
methods to QCD can be motivated from the empirical
evidence~\cite{Deur:2008rf} and theoretical
arguments~\cite{Brodsky:2008be} that the QCD coupling $\alpha_s(Q^2)
$ has an infrared fixed point at low $Q^2.$ 

\begin{figure}[!]
\begin{center}
 \includegraphics[width=15.0cm]{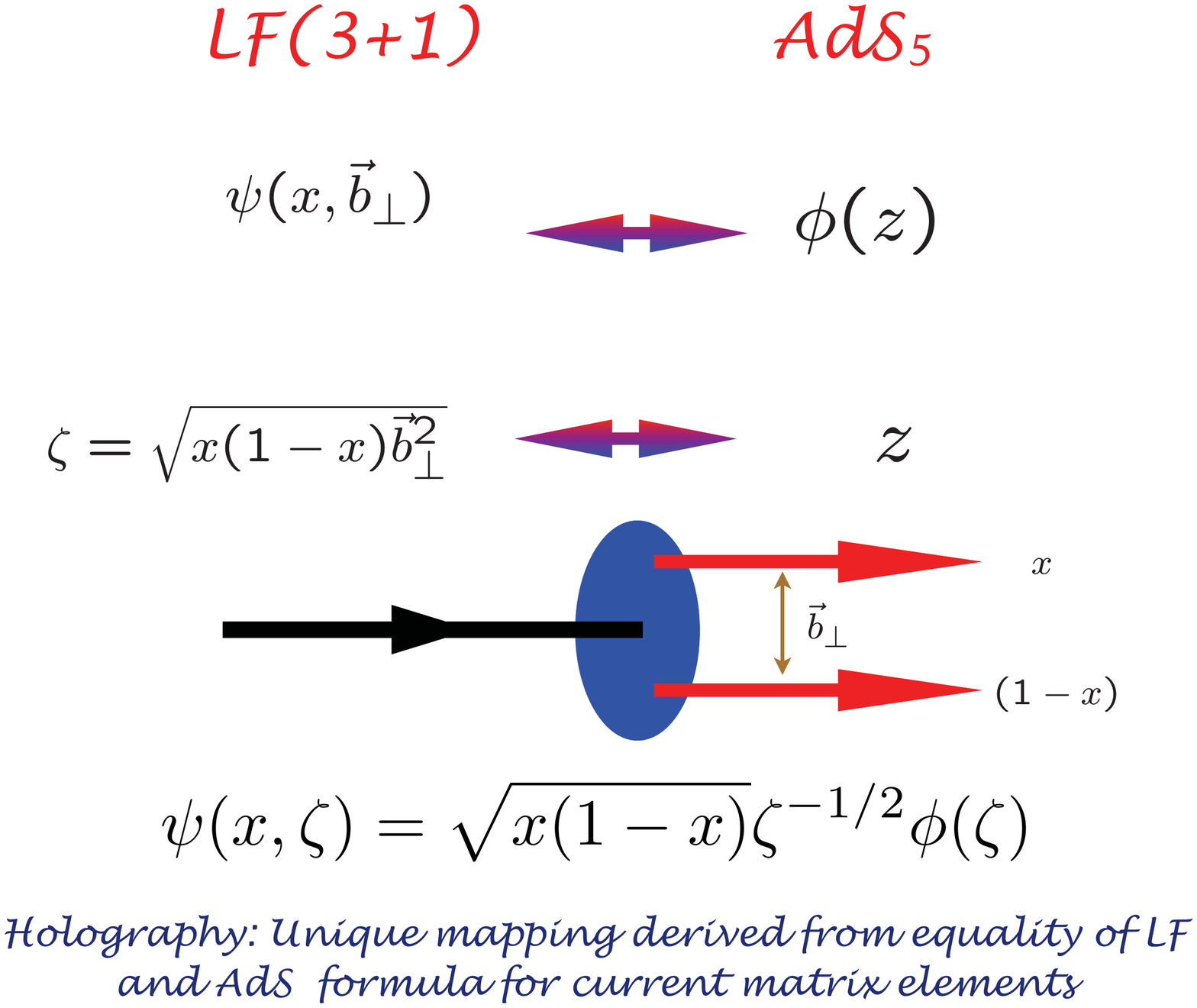}
 \end{center}
 \vspace{-40pt}
 \caption{Light-front holography for meson wavefunctions.  This  mapping is derived from the equality of the LF and AdS  formulae for current matrix elements.}
\label{figholog}  
\end{figure}

The AdS/CFT
correspondence has led to insights into the confining dynamics of
QCD and the analytic form of hadronic light-front wavefunctions. As
Guy de Teramond and I have shown recently, there is a remarkable mapping between the
description of hadronic modes in AdS space and the Hamiltonian
formulation of QCD in physical space-time quantized on the
light-front. This procedure allows string modes $\Phi(z)$ in the AdS
holographic variable $z$ to be precisely mapped to the light-front
wave functions  of hadrons in physical space-time in terms of a
specific light-front variable $\zeta$ which measures the separation
of the quark and gluonic constituents within the hadron. The
coordinate $\zeta$ also specifies the light-front (LF) kinetic
energy and invariant mass of constituents.  See fig.\ref{figholog}.   This mapping was
originally obtained by matching the expression for electromagnetic
current matrix elements in AdS space with the corresponding
expression for the current matrix element using light-front theory
in physical space time~\cite{Brodsky:2006uqa}. More recently we have
shown that one obtains the identical holographic mapping using the
matrix elements of the energy-momentum tensor~\cite{Brodsky:2008pf},
thus providing an important consistency test and verification of
holographic mapping from AdS to physical observables defined on the
light front.

We have also demonstrated~\cite{deTeramond:2008ht}  the equivalence of Anti-de Sitter (AdS) five-dimensional background space  representation with light-front (LF) quantized theory in physical 3+1 space-time. The result is a single-variable light-front Schr\"odinger equation which determines the eigenspectrum and the light-front 
wavefunctions of hadrons for general spin and orbital angular momentum.  In the case of the soft-wall model with massless quarks, the hadronic spectrum has the form of the Nambu string: $M^2 = 4 \kappa^2(n+ L+ S/2)$ in agreement with
experimental results~\cite{Klempt:2007cp}. The scale  $\kappa$ is determined by the soft-wall dilaton metric, $n$ is the principal quantum number, $L$ characterizes the {\it minimum } internal orbital angular momentum of the state and $S$ is the total quark spin.   The pion with $n=0, \, L=0, \, S=0$ is massless, consistent with chiral symmetry.

In general, the hadronic eigensolutions of the LF Hamiltonian equation of motion, and consequently of the AdS equations derived in the framework of light front holography~\cite{deTeramond:2008ht}, have components which span  a set of orbital angular momenta $L$.  For example, as we have shown in \cite{Brodsky:2008pg},  the  spin-1/2  eigensolution for the proton has a Dirac two-component spinor structure 
$\Psi_\pm$ where the upper component has $L=0$ (parallel quark and proton spin) and the lower component has $L=1$ (antiparallel quark and both spins).  Both components are eigensolutions with the same hadronic mass,  consistent with chiral symmetry.  
However, the different orbital components of the hadronic eigensolution have different behavior  as one approaches the short distance $z \to 0$ boundary as predicted from the twist of the relevant interpolating operator.  Only the minimum $L$ term survives at the short distance $x^2=0$, $z=0$ boundary, in agreement with the AdS/CFT dictionary. Thus, only the $S$-wave of the proton eigenstate couples to the relevant interpolating hadronic operator  at $z \to 0$;    the $L=1$ component vanishes with one extra power of $z$.  
Indeed, the relative suppression of the proton LFWF  $\Psi_{-}$ at $z\to 0$ leads to the observation that the struck quark in the proton structure function has the same spin alignment as the proton spin in terms of Bjorken scaling variable $x_{bj} \to 1$.  It also leads to a nonzero neutron Dirac form factor and a higher power-law fall-off for the Pauli versus Dirac form factors of the nucleons.
Thus one only needs the $S$-state to identify the nucleon at $z=0$ in the AdS/CFT description. The $P$-wave component of the eigensolution automatically develops as one evolves the eigenstate to $z > 0.$  
This is in close analogy to the solutions of the Dirac-Coulomb equation in QED. The ground-state eigensolution has both S and P waves (upper and lower components). However, only the S wave appears at the origin ($r=0$).

\begin{figure}[!]
\begin{center}
\includegraphics[width=15.0cm]{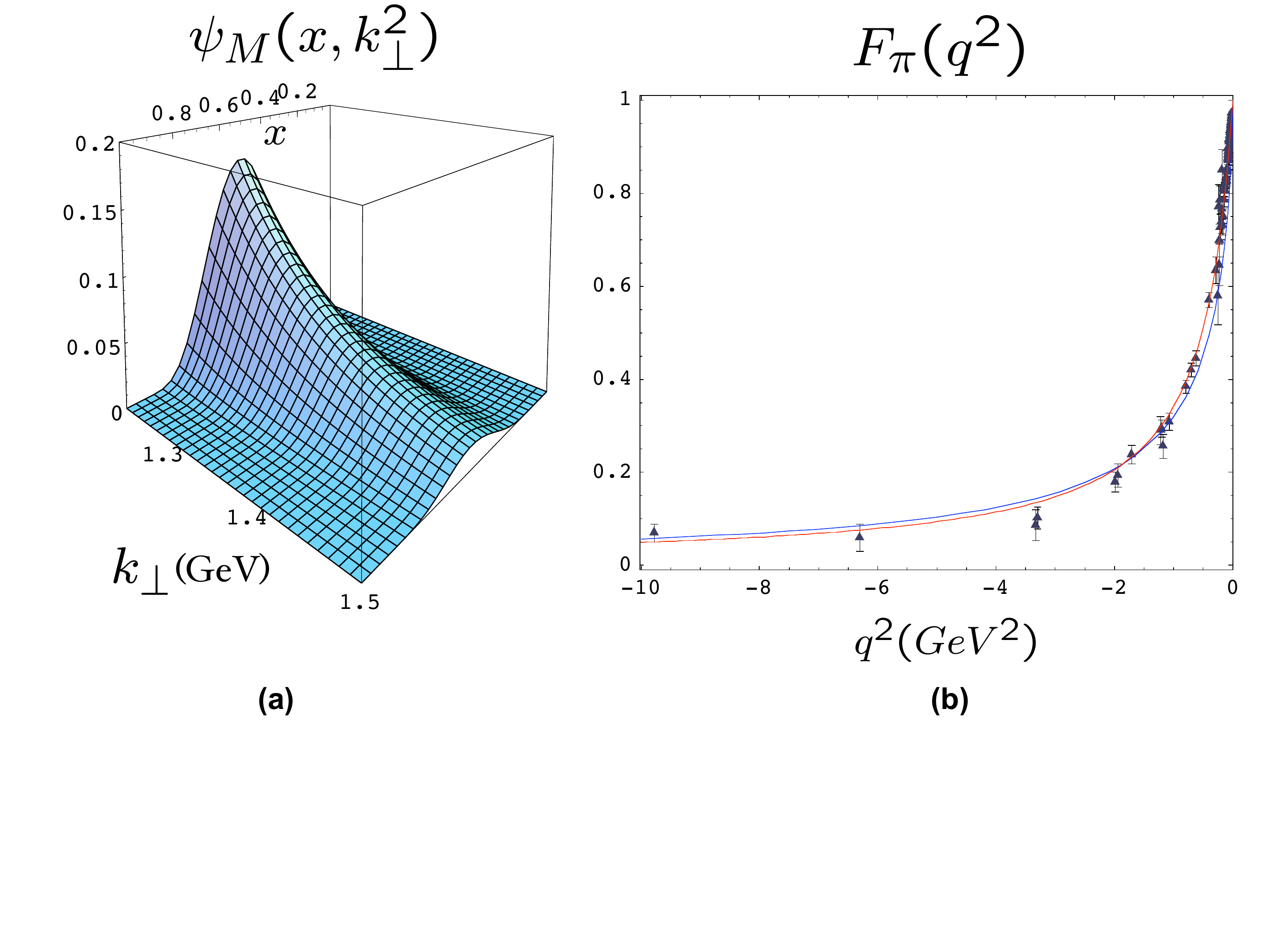}
\end{center}
\vspace{-40pt} 
\caption{(a) Pion light-front wavefunction $\psi_\pi(x, \mbf{b}_\perp$) for the  AdS/QCD  soft wall  ( $\kappa = 0.375$ GeV)  models.  (b) Holographic prediction the space-like pion form factor:  (blue) hard wall ($\Lambda_{QCD} = 0.32$ GeV) and (red) soft wall  ( $\kappa = 0.375$ GeV)  models.}
\label{figff}  
\end{figure}

The pion light-front wavefunction predicted by AdS/QCD and  shown in \ref{figff}(a)  
displays confinement at large interquark
separation and conformal symmetry at short distances, reproducing dimensional counting rules for hard exclusive amplitudes.
The prediction for the spacelike pion form factor for the hard wall and 
soft-wall models is shown in fig.\ref{figff}(b).

The dominance of the minimal $L$ component at $z \to 0$  also yields the dimensional counting rules for form factors at high $Q^2$, as shown by Polchinski and 
Strassler~\cite{Polchinski:2001tt}. 
The leading power fall-off of the hard scattering amplitude as given by dimensional counting rules follows from the conformal scaling of the
underlying hard-scattering amplitude: $T_H \sim 1/Q^{n-4}$, where $n$ is the total number of fields (quarks, leptons, or gauge fields)
participating in the hard scattering~\cite{Brodsky:1974vy,Matveev:1973ra}. Thus the reaction is dominated by subprocesses and Fock states
involving the minimum number of interacting fields.  In the case of $2 \to 2$ scattering processes, this implies $ {d\sigma/ dt}(A B \to C D)
={F_{A B \to C D}(t/s)/ s^{n-2}},$ where $n = N_A + N_B + N_C +N_D$ and $n_H$ is the minimum number of constituents of $H$. The near-constancy
of the effective QCD coupling helps explain the empirical success of dimensional counting rules for the near-conformal power law fall-off of
form factors and fixed angle scaling~\cite{Brodsky:1989pv}.  For example, one sees the onset of perturbative QCD scaling behavior even for
exclusive nuclear amplitudes such as deuteron photodisintegration (Here $n = 1+ 6 + 3 + 3 = 13 .$) $s^{11}{ d\sigma/dt}(\gamma d \to p n) \sim $
constant at fixed CM angle.
The measured deuteron form factor also appears to follow the leading-twist QCD predictions at large momentum transfers in the few GeV
region~\cite{Holt:1990ze,Bochna:1998ca,Rossi:2004qm}.

Conformal symmetry can provide a systematic approximation to QCD in both its nonperturbative and perturbative domains. In the case of
nonperturbative QCD, one can use the AdS/CFT correspondence~\cite{Maldacena:1997re} between Anti-de Sitter space and conformal gauge theories to
obtain an analytically tractable approximation to QCD in the regime where the QCD coupling is large and constant.  Scale-changes in the physical
$3+1$ world can then be represented by studying dynamics in a mathematical fifth dimension with the ${\rm AdS}_5$ metric.

This connection allows one to compute the analytic form~\cite{Brodsky:2006uq,Brodsky:2007pt} of the light-front wavefunctions of mesons and
baryons.  AdS/CFT also provides a non-perturbative derivation of dimensional counting rules for the power-law fall-off of form factors and
exclusive scattering amplitudes at large momentum transfer.
The AdS/CFT approach thus allows one to construct a model of hadrons which has both confinement at large distances and the conformal scaling
properties which reproduce dimensional counting rules for hard exclusive reactions.  The fundamental equation of AdS/CFT has the appearance of a
radial Schr\"odinger Coulomb equation, but it is relativistic, covariant, and analytically tractable.

The deeply virtual Compton amplitudes can be Fourier transformed to $b_\perp$ and $\sigma = x^-P^+/2$ space providing new insights into QCD
distributions~\cite{Burkardt:2005td},\cite{Ji:2003ak},\cite{Brodsky:2006in},\cite{Hoyer:2006xg}. The distributions in the LF direction $\sigma$ typically display
diffraction patterns arising from the interference of the initial and final state LFWFs ~\cite{Brodsky:2006in,Brodsky:2006ku}.  All of these processes  can provide a detailed test of  the AdS/CFT LFWFs predictions.

It is interesting to note that the pion distribution amplitude predicted by AdS/CFT at the hadronic scale is $\phi_\pi(x, Q _0) = {(4/ \sqrt 3
\pi)} f_\pi \sqrt{x(1-x)}$ from both the harmonic oscillator and truncated space models is quite different than the asymptotic distribution
amplitude predicted from the PQCD evolution~\cite{Lepage:1979zb} of the pion distribution amplitude: $\phi_\pi(x,Q \to \infty)= \sqrt 3  f_\pi
x(1-x) $. The broader shape of the AdS/CFT pion distribution increases the magnitude of the leading-twist perturbative QCD prediction for the
pion form factor by a factor of $16/9$ compared to the prediction based on the asymptotic form, bringing the PQCD prediction close to the
empirical pion form factor~\cite{Choi:2006ha}.  An important test of the shape of the pion distribution amplitude is the angular dependence of $\gamma \gamma \to \pi^0 \pi^0.$

Hadron form factors can be directly predicted from the overlap integrals in AdS space or
equivalently by using the Drell-Yan-West formula in physical space-time.   The form factor at high $Q^2$ receives contributions from small
$\zeta \sim {1/ Q}$, corresponding to small $\vec b_\perp$ and  $1-x$ .

\begin{figure}[htb]
\centering
 \includegraphics[width=15.0cm]{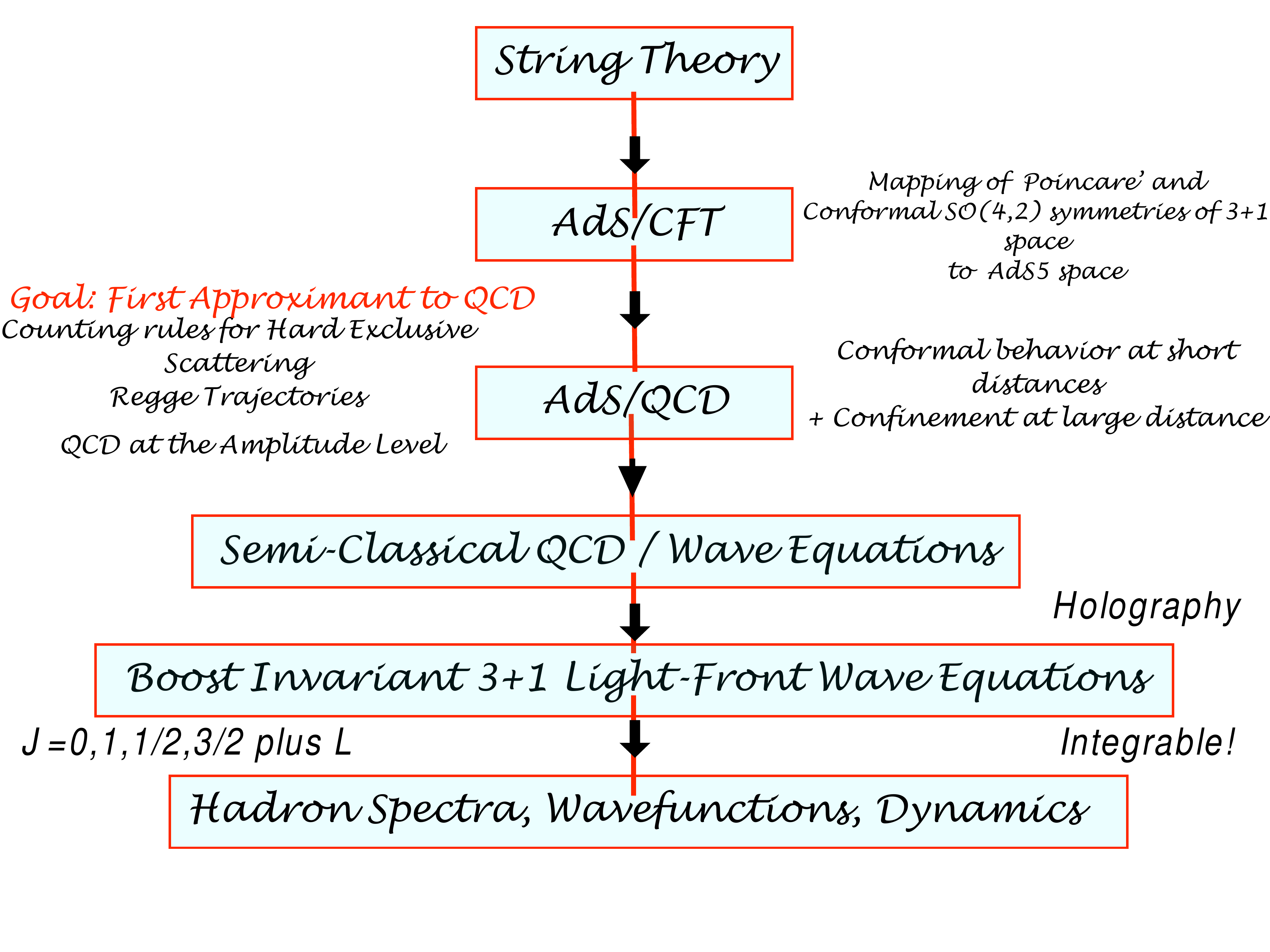}
 \caption{The logistics of AdS/CFT which leads to an analytic first approximation to QCD in its conformal window.}
 \label{figNew25}
 \end{figure}

It is also important to note that the relativistic light-front equation and its eigensolutions can be derived  starting from the light-front Hamiltonian formalism, independent of AdS space considerations~\cite{deTeramond:2008ht}. A dual 5-dimensional background description emerges from LF QCD, so that the powerful geometrical methods from string theory can be used in the description of strongly coupled QCD. The AdS/CFT approach thus provides a viable, analytic  first approximation to QCD. In principle, the model can be systematically improved, for
example by using the AdS/CFT eigensolutions as a basis for diagonalizing the full QCD Hamiltonian. An outline of the AdS/QCD program is shown in
fig.\ref{figNew25}.

The phenomenology of the AdS/CFT model is just beginning, but it can be anticipated that it will have many applications to LHC phenomena. For
example, the model LFWFs provide a basis for understanding hadron structure functions and fragmentation functions at the amplitude level; the
same wavefunctions can describe hadron formation from the coalescence of co-moving quarks.  The spin correlations which underly single and
double spin correlations are also described by the AdS/CFT eigensolutions.  The AdS/CFT hadronic wavefunctions provide predictions for the
generalized parton distributions and weak decay amplitudes from first principles.  In addition, a prediction from AdS/CFT for the proton LFWF
would allow one to compute the higher-twist direct subprocesses such as $ u u \to p \bar d $ which could control proton production in inclusive
reactions at large transverse momenta from first principles. The amplitudes relevant to diffractive reactions could also be computed. We also
anticipate that the extension of the AdS/CFT formalism to heavy quarks will allow a great variety of heavy hadron phenomena to be analyzed from
first principles.

\section{Hadronization at the Amplitude Level}

\begin{figure}[!]
 \begin{center}
\includegraphics[width=15.0cm]{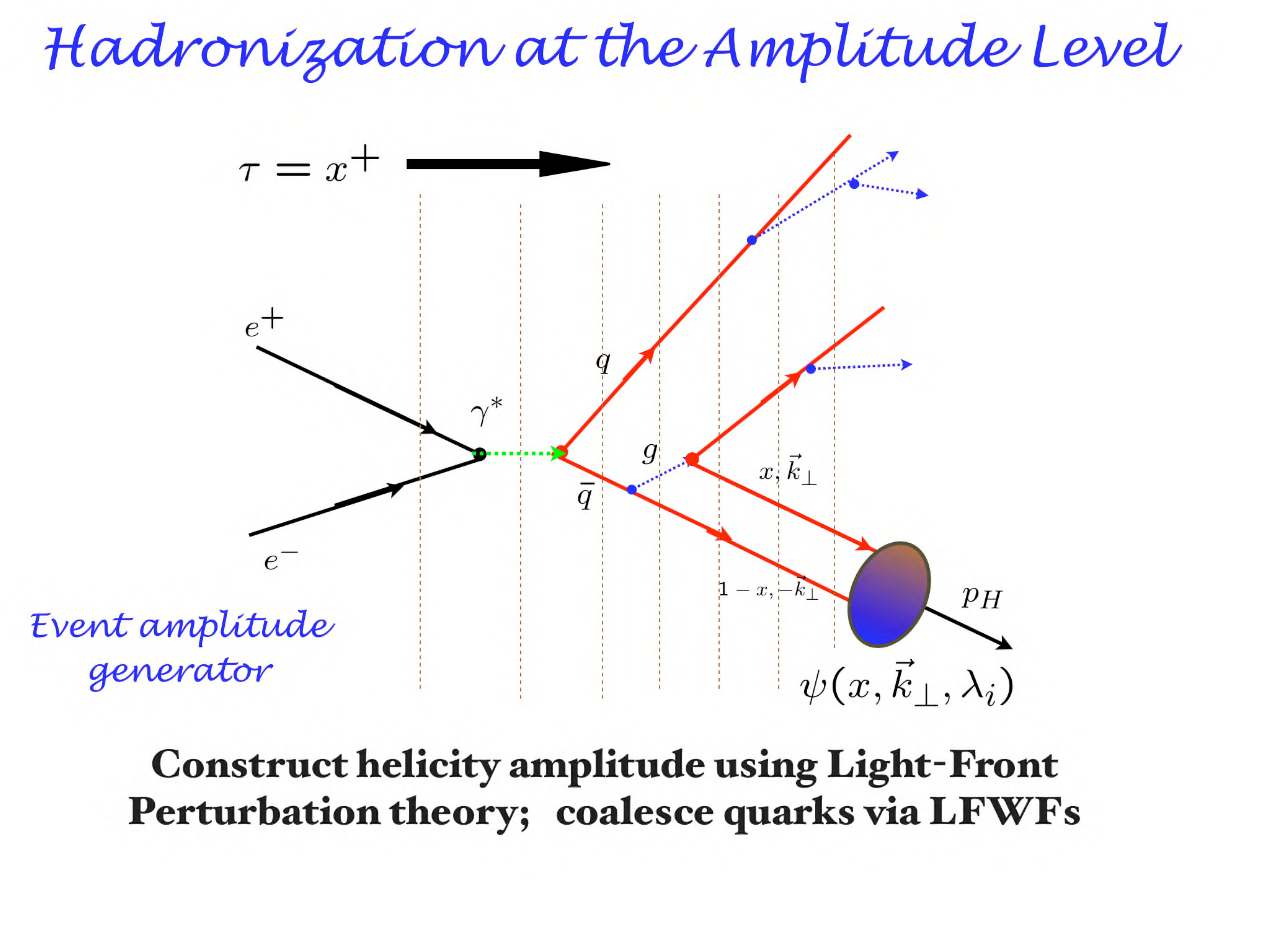}
\end{center}
  \caption{Illustration of an event amplitude generator for $e^+ e^- \to \gamma^* \to X$ for 
  hadronization processes at the amplitude level. Capture occurs if
  $\zeta^2 = x(1-x) \mbf{b}_\perp^2 > 1/ \Lambda_{\rm QCD}^2$
   in the AdS/QCD hard-wall model of confinement;  i.e., if
  $\mathcal{M}^2 = \mbf{k}_\perp^2/x(1-x) < \Lambda_{\rm QCD}^2$.}
\label{figNew18}  
\end{figure}

The conversion of quark and gluon partons is usually discussed in terms  of on-shell hard-scattering cross sections convoluted with {\it ad hoc} probability distributions. 
The LF Hamiltonian formulation of quantum field theory provides a natural formalism to compute 
hadronization at the amplitude level~\cite{Brodsky:2008tk}.  In this case one uses light-front time-ordered perturbation theory for the QCD light-front Hamiltonian to generate the off-shell  quark and gluon T-matrix helicity amplitude  using the LF generalization of the Lippmann-Schwinger formalism:
\begin{equation}
T ^{LF}= 
{H^{LF}_I } + 
{H^{LF}_I }{1 \over {\cal M}^2_{\rm Initial} - {\cal M}^2_{\rm intermediate} + i \epsilon} {H^{LF}_I }  
+ \cdots 
\end{equation}
Here   ${\cal M}^2_{\rm intermediate}  = \sum^N_{i=1} {(\mbf{k}^2_{\perp i} + m^2_i )/x_i}$ is the invariant mass squared of the intermediate state and ${H^{LF}_I }$ is the set of interactions of the QCD LF Hamiltonian in the ghost-free light-cone gauge~\cite{Brodsky:1997de}.
The $T^{LF}$ matrix elements are evaluated between the out and in eigenstates of $H^{QCD}_{LF}.~$ The LFWFs of AdS/QCD can be used as the interpolating amplitudes between the off-shell quark and gluons and the bound-state hadrons.
Specifically,
if at any stage a set of  color-singlet partons has  light-front kinetic energy 
$\sum_i {\mbf{k}^2_{\perp i}/ x_i} < \Lambda^2_{\rm QCD}$, then one coalesces the virtual partons into a hadron state using the AdS/QCD LFWFs.   This provides a specific scheme for determining the factorization scale which  matches perturbative and nonperturbative physics.
The event amplitude generator is illustrated for $e^+ e^- \to \gamma^* \to X$ in fig.\ref{figNew18}.

As I discussed above, the AdS/CFT correspondence between Anti-de Sitter space and conformal gauge theories provides an analytically tractable approximation to
QCD in the regime where the QCD coupling is large and constant.  In particular, there is an exact correspondence between the fifth-dimension
coordinate $z$ of AdS space  and a specific impact variable $\zeta$ which measures the separation of the quark constituents within the hadron in
ordinary space-time. This connection allows one to compute~\cite{Brodsky:2006uq,Grigoryan:2007my} the analytic form of the frame-independent light-front wavefunctions of mesons and
baryons, the fundamental entities which encode hadron properties and allow the computation of exclusive scattering amplitudes.  This opens the possibility of computing hadronization at the amplitude level by convoluting the off-shell hard scattering amplitude with the AdS/QCD light-front wavefunctions.

This scheme has a number of  important computational advantages:

(a) Since propagation in LF Hamiltonian theory only proceeds as $\tau$ increases, all particles  propagate as forward-moving partons with $k^+_i \ge 0$.  There are thus relatively few contributing
 $\tau-$ordered diagrams.

(b) The computer implementation can be highly efficient: an amplitude of order $g^n$ for a given process only needs to be computed once.  In fact, each non-interacting cluster within $T^{LF}$ has a numerator which is process independent; only the LF denominators depend on the context of the process.

(c) Each amplitude can be renormalized using the ``alternate denominator'' counterterm method~\cite{Brodsky:1973kb}, rendering all amplitudes UV finite.

(d) The renormalization scale in a given renormalization scheme  can be determined for each skeleton graph even if there are multiple physical scales.

(e) The $T^{LF}$ matrix computation allows for the effects of initial and final state interactions of the active and spectator partons. This allows for novel leading-twist phenomena such as diffractive DIS, the Sivers spin asymmetry and the breakdown of the PQCD Lam-Tung relation in Drell-Yan processes.

(f)  ERBL and DGLAP evolution are naturally incorporated, including the quenching of  DGLAP evolution  at large $x_i$ where the partons are far off-shell.

(g) Color confinement can be incorporated at every stage by limiting the maximum wavelength of the propagating quark and gluons.

This method retains the quantum mechanical information in hadronic production amplitudes which underlie Bose-Einstein correlations and other aspects of the spin-statistics theorem.
Thus Einstein-Podolsky-Rosen quantum theory correlations are maintained, even between far-separated hadrons and  clusters ~\cite{Abelev:2008ew}.
A similar off-shell T-matrix approach was used to predict antihydrogen formation from virtual positron--antiproton states produced in $\bar p A$  collisions~\cite{Munger:1993kq}.

\section{Setting the Renormalization Scale in Perturbative QCD}

Precise quantitative predictions of QCD are necessary to understand the backgrounds to new beyond-the-Standard-Model phenomena at the LHC . Thus
it is important to eliminate as best as possible all uncertainties in QCD predictions, including the elimination of renormalization scale and
scheme ambiguities.

It is commonly believed that the renormalization scale entering the QCD coupling is an arbitrary parameter in perturbative QCD; in fact,
just as in Abelian theory, the renormalization scale is a physical quantity, representing the summation of QCD vacuum polarization contributions
to the gluon propagator in the skeleton expansion~\cite{Brodsky:1982gc,Grunberg:1991ac}.  In general, multiple renormalization scales appear in a pQCD
expression whenever multiple invariants appear in the reaction.   An important example involving heavy quark production near threshold is discussed in ref.\cite{Brodsky:1995ds}.  In this case one sees that the scale entering the QCD analog of the Sommerfeld-Schwinger-Sakharov correction is of order of $v^2 s,$ where $v$ is the heavy quark velocity in the pair rest frame, not the invariant pair mass $s$.

It should be emphasized that the  renormalization scale is {\it not arbitrary} in gauge theories.  For example in QED, the renormalization scale
in the usual Gell Mann-Low scheme is  exactly the photon virtuality: $\mu^2_R = k^2$. This scale sums all vacuum polarization corrections into
the dressed photon propagator of a given skeleton graph. The resulting analytic QED running coupling has dispersive cuts set correctly set at
the physical thresholds for lepton pair production $k^2= 4m^2_L$. (In ${\overline MS}$ scheme, the renormalization scales are displaced  to
$e^{-5/3} k^2$.) The renormalization scale is similarly unambiguous in QCD: the cuts due to quark loops  in the dressed gluon propagator appear
at the physical quark thresholds.  Equivalently, one can use the scheme-independent BLM
procedure~\cite{Brodsky:1982gc,Brodsky:1994eh,Grunberg:1991ac} to eliminate the appearance of the $\beta$-function in the perturbative series.

Of course the {\it initial choice} of the renormalization  scale is completely arbitrary, and one can study the dependence of a perturbative
expansion on the initial scale using the usual renormalization group evolution equations.  This procedure exposes the $\beta-$dependent terms in
the PQCD expression. Eliminating the $\beta$-dependent terms then leads to a unique, physical, renormalization scale for any choice of
renormalization scheme.  In effect, one identifies the series for the corresponding conformal theory where the $\beta-$ function is zero;  the
conformal expression serves as a template~\cite{Brodsky:1999gm} for perturbative QCD expansions; the nonzero QCD $\beta$-function can then be
systematically incorporated into the scale of the running coupling~\cite{Brodsky:1994eh,Brodsky:1995tb,Brodsky:2000cr}. This leads to fixing of
the physical renormalization scale as well as commensurate scale relations which relate observables to each other without scale or scheme
ambiguity~\cite{Brodsky:1982gc}.

As an example, consider Higgs production $ p p \to H X$ calculated via $g g \to H$ fusion. The physical renormalization scale for the running
QCD couplings for  this subprocess in the pinch scheme are  the two gluon virtualities, not the Higgs mass.  The resulting values for the
renormalization scales parallel the two-photon process in QED:  $e e \to e e H$ where only vacuum polarization corrections determine the scale;
i.e., the renormalization scales are set by the photon virtualities. An interesting consequence is the prediction that the QCD coupling is
evaluated at the minimal scale of the gluon virtualities  if the Higgs is measured at $\vec p^H_T =0$.

In a physical renormalization scheme~\cite{Grunberg:1982fw}, gauge couplings are defined directly in terms of physical observables. Such
effective charges are analytic functions of the physical scales and their mass thresholds  have the correct threshold
dependence~\cite{Brodsky:1998mf,Binger:2003by} consistent with unitarity. As in QED, heavy particles contribute to physical predictions even at
energies below their threshold. This is in contrast to renormalization schemes such as $\bar{MS}$ where mass thresholds are treated as step
functions.  In the case of supersymmetric grand unification, one finds a number of qualitative differences and improvements in precision over
conventional approaches~\cite{Binger:2003by}. The analytic threshold corrections can be important in making the measured values of the gauge
couplings consistent with unification.

Relations between observables  have no scale ambiguity and are independent of the choice of the intermediate renormalization
scheme~\cite{Brodsky:1994eh};  this is the transitivity property of the renormalization group. The results, called commensurate scale relations,
are consistent~\cite{Brodsky:1992pq} with the renormalization group~\cite{Stueckelberg:1953dz} and the analytic connection  of QCD to Abelian
theory at $N_C\to 0$~\cite{Brodsky:1997jk}.  A important example is the generalized Crewther relation~\cite{Brodsky:1995tb}. One finds a
renormalization-scheme invariant relation between the coefficient function for the Bjorken sum rule for polarized deep inelastic scattering and
the $R$-ratio for the $e^+e^-$ annihilation cross section. This relation provides a generalization of the Crewther relation to non-conformally
invariant gauge theories. The derived relations allow one to calculate unambiguously without renormalization scale or scheme ambiguity the
effective charges of the polarized Bjorken and the Gross-Llewellen Smith sum rules from the experimental value for the effective charge
associated with $R$-ratio. Present data are consistent with the generalized Crewther relations, but measurements at higher precision and
energies are needed to decisively test these fundamental relations in QCD.

Binger and I ~\cite{Binger:2006sj} have analyzed the behavior of the thirteen nonzero form factors contributing to the
gauge-invariant three-gluon vertex at one-loop, an analysis which is important for setting the renormalization scale for heavy quark production
and other PQCD processes. Supersymmetric relations between scalar, quark, and gluon loops contributions to the triangle diagram lead to a simple
presentation of the results for a general non-Abelian gauge theories.  Only the gluon contribution to the form factors is needed since the
massless quark and scalar contributions are inferred from the homogeneous relation $F_G+4F_Q+(10-d)F_S=0$ and the sums $\Sigma_{QG}(F) \equiv
{(d-2)/ 2}F_Q + F_G$ which are given for each form factor $F$. The extension to the case of internal masses leads to the modified sum rule
$F_{MG}+4F_{MQ}+(9-d)F_{MS}=0$. The phenomenology of the three-gluon vertex is largely determined by the form factor multiplying the three-level
tensor. One can define a three-scale effective scale $Q^2_{eff}(p^2_a,p^2_b,p^2_c)$ as a function of the three external virtualities which
provides a natural extension of BLM scale setting \cite{Brodsky:1982gc} to the three-gluon vertex. Physical momentum scales thus set the scale
of the coupling. 

\section{Summary}
I have discussed several novel phenomenological features of QCD in high transverse momentum reactions at RHIC, such as the ridge,  the baryon anomaly,  and the breakdown of leading-twist scaling. 
I suggest that the ridge is a consequence of the semihard gluons associated with the trigger-biased DLGLAP evolution of the initial-state quark and gluon distributions,  imprinted on the nuclear medium. The
presence of direct higher-twist processes, where a proton is produced directly in the hard subprocess, can explain  the large proton-to-pion ratio in
high-centrality heavy-ion collisions.  Direct hadronic processes can also account for the deviation from leading-twist PQCD scaling at fixed $x_T= 2 p_T/\sqrt s$  as well as explain the change in the effective power $n_{eff}$ for proton production with centrality.

Initial- and
final-state interactions from gluon-exchange, normally neglected in the parton model, have a profound effect in QCD hard-scattering reactions,
leading to leading-twist single-spin asymmetries, diffractive deep inelastic scattering, diffractive hard hadronic reactions, the breakdown of
the Lam Tung relation in Drell-Yan reactions, nuclear shadowing and non-universal antishadowing---leading-twist physics not incorporated in
the light-front wavefunctions of the target computed in isolation. 

I have discussed other aspects of quantum correlations in heavy ion collisions, such as tests of hidden color in nuclear wavefunctions, the use of
diffraction to materialize the Fock states of a hadronic projectile and to test QCD color transparency. The  consequences of color-octet  intrinsic heavy quarks such as high-$xF$  Higgs production were also reviewed.  A model for the multiple scattering and early thermalization of the quark-gluon medium was outlined.  I also discussed how the AdS/CFT correspondence between Anti-de Sitter space and conformal
gauge theories provides an exact correspondence  for computing the
analytic form of the frame-independent light-front wavefunctions of mesons and baryons, providing a method to compute hadronization at the amplitude level.  Finally, the elimination of the renormalization scale ambiguity in PQCD calculations was discussed.

\begin{acknowledgments}
I thank the organizers of the  4th International Workshop on High-$p_T$ physics at the LHC,
especially Professor Jan Rak,  for their invitation to this workshop.  I also thank Michael Tannenbaum, Barbara Jacek, and my collaborators, especially Francois Arleo, Patrick Aurenche, Anne Sickles, Dae Sung Hwang, Michael Binger, Ivan Schmidt,  Boris Kopeliovich, Jorg Raufeisen, Hans-Jurgen Pirner, Jack Soffer, Susan Gardner, Fred Goldhaber, Paul Hoyer, and Guy de Teramond, for many helpful discussions. This work was supported in
part by the Department of Energy, contract No. DE-AC02-76SF00515.
\end{acknowledgments}

\end{document}